\documentclass[a4paper,11pt,headings=big,DIV=12]{article}
\pdfoutput=1 
\usepackage{graphicx}% Include figure files
\usepackage{dcolumn}% Align table columns on decimal point
\usepackage{bm}% bold math 
\usepackage{amssymb}
\usepackage{graphicx} 
\usepackage{amsbsy}
\usepackage{amsmath} 
\usepackage{slashed}
\usepackage{amsfonts}
\usepackage{comment}
\usepackage{braket}
\usepackage{tensor}
\usepackage[scriptsize]{subfigure}
\usepackage{float}
\usepackage{enumerate}
\usepackage{multirow}
\usepackage{yfonts}

\usepackage{jheppub}

\usepackage[utf8]{inputenc}

\usepackage{listings}
\usepackage{color}

\definecolor{mygreen}{rgb}{0,0.6,0}
\definecolor{mygray}{rgb}{0.5,0.5,0.5}
\definecolor{mymauve}{rgb}{0.58,0,0.82}

\definecolor{darkWhite}{rgb}{0.94,0.94,0.94}

\newcommand{\DP}{\textfrak{D}}
\newcommand{\dd} {\mathrm{d}}
\newcommand{\tr} {\mathrm{tr}}
\newcommand{\Tr} {\mathrm{Tr}}

\newcommand{\Lagr}{\mathcal{L}}

\newcommand{\g}{\sqrt{ g }}

\newcommand{\sD}{\slashed{D}}
\newcommand{\Od}{\mathcal{O}(d-4)}

\DeclareMathOperator{\pa}{\partial}

\newcommand{\vev}[1]{\langle #1 \rangle}

\newcommand{\al}{\alpha}
\newcommand{\be}{\beta}
\newcommand{\ga}{\gamma}

\newcommand{\de}{\delta}

\newcommand{\la}{\lambda}

\newcommand{\eps}{\epsilon}

\newcommand{\sig}{ \sigma}

\newcommand{\om}{\omega}

\newcommand{\EQ}{Eq.~}
\newcommand{\EQs}{Eqs.~}
\newcommand{\TAB}{Tab.~}

\newcommand{\SEC}{Sec.~}
\newcommand{\SECs}{Secs.~}
\newcommand{\APP}{App.~}
\newcommand{\APPs}{Apps.~}

\newcommand{\TEMT}{ \Tud{\rho}{\rho} }
\newcommand{\Tud}[2]{T^{#1}_{\;\; #2}}

\newcommand{\ORD}{{\cal O}}

\newcommand{\RRt}{R \tilde R}

\newcommand{\DM}{ {\cal D}}

\newcommand{\Weff}{{\cal W}}
\newcommand{\Wren}{\Weff_{\text{ren}}}
\newcommand{\Wct}{\Weff_{\text{ct}}}
\newcommand{\Wctodd}{\Weff_{\text{ct,odd}}}

\newcommand{\Tct}{\mathcal{T}_{\text{ct}}}

\newcommand{\fin}{{\text{fin}}}
\newcommand{\Div}{{\text{div}}}

\usepackage{dsfont}

\usepackage{hyperref}

\usepackage{color}
\definecolor{verdes}{cmyk}{0.92,0,0.59,0.4}

\definecolor{Grn}{rgb}{0.1,0.5,0.2}
\definecolor{Blu}{rgb}{0.,0.,0.1}
\definecolor{Red}{rgb}{0.7,0.1,0.1}
\definecolor{SE}{rgb}{0.5,0,0.4}
\definecolor{Tur}{rgb}{0,0.75,0.65}

\usepackage{hyperref} 
\hypersetup{
     colorlinks = true,
     linkcolor = black,
     citecolor = blue,
     anchorcolor = blue,
     filecolor = blue,
     urlcolor = blue,
     bookmarks=true,
     linktocpage =true,
     }

\begin{document}

\title{\boldmath  Gravity-gauge  Anomaly Constraints \\ on the Energy-momentum Tensor     
   }

\author[1]{R\'emy Larue,}
\author[2]{J\'er\'emie Quevillon,}
\author[3]{Roman Zwicky,}

\affiliation[1]{ Laboratoire de Physique Subatomique et de Cosmologie,
	 Universit\'{e} Grenoble-Alpes, CNRS/IN2P3, Grenoble INP, 38000 Grenoble, France}
\affiliation[2]{Laboratoire d’Annecy-le-Vieux de Physique Th\'eorique,
CNRS – USMB, BP 110 Annecy-le-Vieux, 74941 Annecy, France}
\affiliation[4]{Higgs Centre for Theoretical Physics, School of Physics and
Astronomy, The University of Edinburgh, 
Peter Guthrie Tait Road, Edinburgh EH9 3FD, Scotland, UK}
\emailAdd{jeremie.quevillon@lapth.cnrs.fr}
\emailAdd{remy.larue@lpsc.in2p3.fr}
\emailAdd{roman.zwicky@ed.ac.uk}

\abstract{We derive constraints on the four dimensional energy-momentum tensor 
from gravitational and gauge anomalies.  
Our  work can be considered an extension of Duff's analysis \cite{DUFF1977334} to include parity-odd terms 
and explicit symmetry breaking.  The constraints imply the absence of   the parity-odd  $R\tilde R$ and $F\tilde F$ terms, 
for theories whose symmetries are compatible with dimensional regularisation, 
in a model-independent way.
Remarkably, even in the case of explicit symmetry breaking the $\Box R$-anomaly is found to be finite and unambiguous after applying the symmetry constraints. We compute 
  mixed gravity-gauge anomalies at leading order and deduce phenomenological consequences 
  for vector bosons associated with global chiral  symmetries. \\

}

\maketitle

\section{Introduction}

Weyl anomalies in four dimensions have been discovered  fifty years ago \cite{Capper:1974ic}, they 
are connected to many fields of physics \cite{Duff:1993wm} and are still an active field of research. 
The anomaly appears in the divergence of the dilatation current which is given 
by the trace of the energy momentum tensor (TEMT)
\begin{equation}
\label{Eq1Weyl}
g^{\al \be} \vev{T_{\al \be}}  =a\, E +b\,R^2 +c\,W^2 +d\,\Box R+e\,  \RRt  + \, f F^2 + \, h F\tilde F  \;,
\end{equation}
where $E$, $R^2$, $W^2$, $ \RRt$ and $\Box R$ are invariants of the Riemann tensor and $F^2$ and $F\tilde F$ are  the standard field strength tensor invariants (all defined in the main text). 
The coefficients are model-dependent and a non-zero value signals  anomalous breaking of Weyl invariance
in the case where the theory is classically Weyl invariant.

Constraints on the TEMT were worked out at leading order (LO) in a classic paper by Duff within dimensional regularisation \cite{DUFF1977334}.
The main goal of our work is to extend it to include: i) the case of parity-odd terms and ii) explicit symmetry breaking. The parity-odd terms consist of the topological Pontryagin densities $\RRt$ and $F\tilde F$ 
which each contain a single Levi-Civita tensor. If their respective coefficients are real (imaginary) 
then the terms violate  CP (CPT) in addition.  There has been a recent controversy as to whether 
Weyl fermions can give rise to an $\RRt$-term. The group of Bonora has found that there ought to be 
 \cite{Bonora:2014qla,Bonora:2015nqa,Bonora:2017gzz,Bonora:2018ajz,Bonora:2018obr,Bonora:2022izj}
whereas all other investigations have concluded to the contrary \cite{Bastianelli:2016nuf,Bastianelli:2019zrq,Frob:2019dgf,Abdallah:2021eii,Abdallah:2023cdw,Larue:2023tmu}. In our previous work  \cite{Larue:2023tmu} 
we have analysed the issue via the path integral paying attention to the definition of the Weyl determinant.  
In this work we derive the vanishing $e=0$ from the decomposition of the EMT-counterterms  and our results hold 
for all theories whose symmetries are compatible with dimensional regularisation,  which  includes the  spin-3/2 case (and also spin-2)  \cite{Christensen:1978gi,Christensen:1978md,Birrell:1982ix,Perry:1978jj,Critchley:1978ki,Yoneya:1978gq,Critchley:1978kb,Fradkin:1982bd,Fradkin:1982kf,Fradkin:1985am}.
In addition we  consider the parity-even sector with explicit breaking of 
Weyl symmetry.  Constraints are implied by the \emph{finiteness}  of the diffeomorphism (diffeo) and Lorentz anomalies. 
A priori this is not obvious as one might think that their presence invalidates the theory. 
However, this is not the case   since they can, for example, be seen as low energy effective theories of decoupled
 chiral fermions in the context of the Higgs mechanism   \cite{Ball:1988xg,Preskill:1990fr}. 
On a practical level finiteness of the diffeo, Lorentz and gauge anomalies is due to their topological nature \cite{Adler:1969er,Alvarez-Gaume:1983ihn}. In the second part of the paper we  include gauge fields and work out   
diffeo, Lorentz and gauge anomaly constraints which automatically includes mixed anomalies. 
Where possible we find agreement with the literature but we believe that our diffeo-anomaly constraints 
are more complete. Phenomenological implications for vector bosons  associated with   global  chiral symmetries 
are briefly outlined.

The paper is organised as follows.  \SEC\ref{sec:anomalydef} contains definitions  
 and the derivation of  a formula, widely used in the literature, for the Weyl anomaly in the 
 presence of explicit symmetry breaking.
\SEC\ref{sec:gen} is the central part of this work where all possible terms of the  EMT-counterterms are considered
and  constraints due to the diffeo and the Lorentz anomalies are worked out.  
In \SEC\ref{sec:gauge} the analysis is extended to include a gauge sector. 
Finally, in \SEC\ref{sec:mixed} we {explicitly compute the mixed gravity-gauge anomalies for a Weyl fermion}, 
compare to the literature and deduce phenomenological consequences. The paper ends with summary and 
conclusions in \SEC\ref{sec:conc}.  Further derivations and 
 constraints of the EMT
are deferred to 
 \APPs \ref{app:AnomalyDef} and \ref{app:EMTconstraints}  respectively.

\section{The Weyl Anomaly and (no-)explicit Symmetry Breaking} 
\label{sec:anomalydef}
 
 In \SEC\ref{sec:defs} some basic definitions are given before  discussing the Weyl anomaly 
 with and without explicit symmetry breaking in \SECs \ref{sec:anomalyCFT} and  \ref{sec:anomalyEX}  respectively.

\subsection{Preliminary definitions }
\label{sec:defs}

Let us consider a theory where  quantum fields are collectively denoted by $\phi$.  The metric 
$g_{\mu\nu}$ and the gauge field $A_\mu$ are considered as external fields, with  
Weyl transformations
\begin{equation}
\delta^W_\sig g_{\mu\nu}=2\sig(x) g_{\mu\nu}\;,\quad\quad \delta^W_\sig A_\mu=0\;.
\end{equation}
In a space with Euclidean signature\footnote{Regarding the mapping between Minkowski and Euclidian signatures for spinors, see e.g \cite{Mehta:1986mi,Kupsch:1988sb,vanNieuwenhuizen:1996tv}.} the quantum effective action $\Weff$ is given by
\begin{equation}
\Weff[g_{\mu\nu},A_\mu]=-\log \int \DP\phi \, e^{-S[g_{\mu\nu},A_\mu,\phi]}\; .
\end{equation}
The renormalised effective action  is defined as the sum of the bare and the counterterm effective action 
\begin{equation}
\label{eq:Wren}
\Wren (d)=\Weff(d)+\Wct (d)\; ,
\end{equation}
which is finite when the regulator is removed  ($d \to 4$ in dimensional regularisation with $d = 4 - 2\eps$ adopted throughout). 
The separate pieces  $\Weff(d)$ and $\Wct(d)$ contain $\frac{1}{d-4}$-terms at leading order (LO)  \cite{DUFF1977334}\footnote{For the time being, we have omitted 
the gauge field term $\tr\, F^2$.}
\begin{equation}
\label{eq:Wctint}
\Wct(d)=\frac{1}{d-4}\int\dd^d x\, \g \,\left(a\,E(d)+b\,R^2(d) + c\, C(d)\right)\;,
\end{equation}
and the integrand is a  local polynomial in the background fields.  Note that no  $\Box R$-term is 
included at the level of the action since it is a total derivative. The possibility of adding a $P$ and $CP$-violating 
term $\RRt$ will be discussed in \SEC\ref{sec:gen}.
For a classically Weyl invariant theory, the Weyl consistency conditions imply 
 $b=0$   \cite{DUFF1977334}.  The remaining two terms $E$ and $C$ are given by
 \begin{alignat}{2}\label{eq:E,C}
&E(d)&\;=\;&R_{\mu\nu\rho\sig}R^{\mu\nu\rho\sig}-4R_{\mu\nu}R^{\mu\nu}+R^2\;, \nonumber \\
&C(d)&\;=\;&R_{\mu\nu\rho\sig}R^{\mu\nu\rho\sig}-2 R_{\mu\nu}R^{\mu\nu}+\frac{1}{3}R^2\;,
\end{alignat}
where $E(4)$ reduces to the topological Euler density and $C(4)$ to the Weyl tensor squared.
The curvature invariants in non-integer dimension are understood as power series in the metric \cite{DESER197645}.
The (unrenormalised) EMT and its vacuum expectation value (VEV) are given by the metric variation of  
the (effective) action 
\begin{equation}
\label{eq:TEMT}
T^{\mu\nu}=\frac{2}{\g}\frac{\delta S}{\delta g_{\mu\nu}}
\;,\quad\quad\langle T^{\mu\nu} \rangle=\frac{2}{\g}\frac{\delta \Weff}{\delta g_{\mu\nu}}\;.
\end{equation}
Above  $g$ denotes the metric determinant.   The Weyl variation of the action yields the (classical) TEMT when $\phi$ is on-shell
\begin{equation}
\delta^W_\sig S=\int\dd^4x\,\g\,\sig\, \TEMT\;.
\end{equation}
In the literature the Weyl anomaly is often defined as 
\cite{Duff:1993wm,Casarin:2018odz,Duff:2020dqb,Casarin:2021fgd}\footnote{The superscript denotes the dimension in which the metric lives and we have $g^{(d)}_{\mu\nu}=g^{(4)}_{\mu\nu}+\Od$. 
When  the superscript is omitted $d$-dimensional tensors  are assumed.}$^,$\footnote{Note that whereas  for $d=4-2 \eps$ one has 
$g^{(d)}_{\mu\nu}\langle T^{\mu\nu}\rangle = \langle g^{(d)}_{\mu\nu} T^{\mu\nu}\rangle=\vev{\TEMT}$, this no longer holds for $g^{(4)}_{\mu\nu}$.}
\begin{equation}
\label{eq:AnomalyDef}
\mathcal{A}_{\mathrm{Weyl}}= g^{(4)}_{\mu\nu}\langle T^{\mu\nu}\rangle-\langle g^{(d)}_{\mu\nu}T^{\mu\nu}\rangle \;,
\end{equation}
within dimensional regularisation. 
The objective of this section is to formally derive this formula   accounting for explicit symmetry breaking -- 
an  aspect seemingly absent in the 
literature.\footnote{A new definition of the anomaly  beyond dimensional regularisation has 
   very recently been proposed in \cite{Ferrero:2023unz}.}

\subsection{Weyl anomaly with classical  Weyl invariance }
\label{sec:anomalyCFT}

Let us  first consider the case of a theory which is classically Weyl invariant: $\delta^W_\sigma S =0$.  
To define the anomaly from the unrenormalised EMT bears an ambiguity in that the Weyl variation can 
be performed in four or $d$ dimension: $g^{(4)}_{\mu\nu} \vev{T^{\mu\nu}}\neq \vev{\tensor{T}{^\mu_\mu}}  $. This ambiguity is absent if one directly defines the Weyl anomaly from 
the renormalised effective action 
\begin{equation}
\label{eq:anomalyWrenCFT}
\int\dd^4x\,\g\,\sigma\mathcal{A}_{\mathrm{Weyl}} \equiv \delta^W_\sig \lim_{d\to4} \Wren (d)=\lim_{d\to4}\delta^W_\sigma \Wren(d)\;,
\end{equation}
since the finiteness of $\Wren$ guarantees  that the Weyl variation and the $d\to 4$  limit  commute. 
Applying the first definition results in\footnote{This expression of the anomaly was found in \cite{DESER197645} in $d=2-\eps$ using a Taylor expansion in $\eps$. This assumes continuity of the tensors which may not always hold, as we will discuss in \SEC\ref{sec:gen}.}
\begin{equation}\label{eq:anomalyWCFT}
 \delta^W_\sig \lim_{d\to4} \Wren (d)  \equiv \int\dd^4x\, \sig \, g^{(4)}_{\mu\nu}\left[\frac{\delta \Wren(d)}{\delta g_{\mu\nu}}\right]_{d=4}= \int\dd^4x\,\sig \g \, g^{(4)}_{\mu\nu}\langle T^{\mu\nu}\rangle\;,
\end{equation}
since classical Weyl invariance implies \cite{DUFF1977334}
\begin{equation}
\label{eq:1}
\left[ g^{(4)}_{\mu\nu}\frac{\delta \Wct}{\delta g_{\mu\nu}} \right]_{d=4}=0 \;.
\end{equation}
If we apply the second definition we obtain,  the expression used in \cite{DUFF1977334},
\begin{equation}
\label{eq:CFT,A,Wct}
\lim_{d\to4}\delta^W_\sigma \Wren(d) \equiv \int\dd^4x\, \sig \left[g^{(d)}_{\mu\nu}\frac{\delta \Wren(d)}{\delta g_{\mu\nu}}\right]_{d=4}=\int\dd^4x\, \sig \left[g^{(d)}_{\mu\nu}\frac{\delta \Wct(d)}{\delta g_{\mu\nu}}\right]_{d=4}\;,
\end{equation}
since 
\begin{equation}
\label{eq:2}
\left[g^{(d)}_{\mu\nu}\vev{T^{\mu\nu}} \right]_{d=4} = \left[\vev{\tensor{T}{^\mu_\mu}}\right]_{d=4} =0 \;,
\end{equation}
by classical Weyl invariance, with more details in  \APP\ref{app:AnomalyDef}.   
This demonstrates formula \eqref{eq:AnomalyDef} for a classically Weyl invariant theory 
and clarifies why the Weyl anomaly  can be obtained in different ways.
Similar reasonings have been applied in \cite{Casarin:2021fgd} (cf. \SEC 1.2.3.) to show that the anomaly can be obtained
in both ways.   

Note that the form of the anomaly is constrained by the Wess-Zumino consistency conditions \cite{Wess:1971yu} in a classically Weyl invariant theory, since it is given by the variation of a functional \eqref{eq:anomalyWrenCFT}.\footnote{The Wess-Zumino consistency conditions for the Weyl transformation have been investigated  for a Weyl invariant theory in full generality some time ago \cite{Osborn1991WeylCC}.} This aspect will be exploited in \SEC\ref{sec:WZCC}.

Anomalies are obtained by the variation of the bare effective action $\Weff$ 
and since they are finite, the variation of $\Wct$  does not enter in general. We will use this invariance of $\Wct$ under the diffeo, Lorentz and gauge transformations later on. For the Weyl anomaly, it manifests itself in \eqref{eq:1}. On the other hand, the Weyl anomaly is known to be related to the renormalisation of the theory (e.g \cite{Osborn1991WeylCC,Fujikawa:2004cx}). This is apparent in the second expression of the Weyl anomaly  \eqref{eq:CFT,A,Wct}, which is finite since the $d$-dimensional variation produces a term proportional to $d-4$.

\subsection{Weyl anomaly without classical  Weyl invariance   }
\label{sec:anomalyEX}

We turn to the case of a theory with explicit Weyl symmetry breaking: $\delta^W_\sigma S\neq0$. In that case, 
formula \eqref{eq:anomalyWrenCFT} is amended\footnote{Note that \eqref{eq:anomalyWren} equals to
   $\left[g^{(4)}_{\mu\nu}\langle T^{\mu\nu}\rangle\right]_{\text{fin}}$ with ``fin" referring to the finite part.} 
\begin{equation}\label{eq:anomalyWren}
\int\dd^4x\,\g\,\sigma\left(\mathcal{A}_{\mathrm{Weyl}}+\mathcal{E}_{\mathrm{Weyl}}\right)=\delta^W_\sig \lim_{d\to4} \Wren(d)=\lim_{d\to4}\delta^W_\sigma \Wren(d)\;,
\end{equation}
to include a term   that parametrises the explicit breaking. 
It is worthwhile to emphasise that both 
 $\mathcal{A}_{\mathrm{Weyl}}$ and $\mathcal{E}_{\mathrm{Weyl}}$  are finite quantities. 
 The results in the previous section, that is \EQs \eqref{eq:1} and \eqref{eq:2},  will not apply as $\Wct$ does not need to consist of 
 Weyl invariant terms.  Applying the two definitions one gets
\begin{alignat}{3}
\label{eq:noCFT,A+E}
&\delta^W_\sig \lim_{d\to4} \Wren(d) &\;=\;&\int\dd^4x\, \sig \Big[ \g g^{(4)}_{\mu\nu}\langle T^{\mu\nu}\rangle &\;+\;&g^{(4)}_{\mu\nu}\frac{\delta \Wct(d)}{\delta g_{\mu\nu}}\Big]_{d=4} \;,  \nonumber\\
&\lim_{d\to4}\delta^W_\sigma \Wren(d) &\;=\;& \int\dd^4x\, \sig \Big[ \g\langle\tensor{T}{^\mu_\mu}\rangle&\;+\;&g^{(d)}_{\mu\nu}\frac{\delta \Wct(d)}{\delta g_{\mu\nu}} \Big]_{d=4}\;,
\end{alignat}
upon  using \EQ\eqref{eq:Wren}. In each line, both terms are separately divergent but finite in their sum.
The main idea is to define $ \mathcal{E}_{\mathrm{Weyl}}$ from the finite part of 
$\vev{ \tensor{T}{^\mu_\mu}}$ 
and,  using the second line of \eqref{eq:noCFT,A+E}, this further implies\footnote{\label{foot:finiteterminWct} Note that we assumed a purely divergent $\Wct$ (i.e minimal subtraction scheme). \eqref{eq:EA} has to be amended if we include a finite piece in $\Wct$, but the anomaly remains unaffected (see \APP\ref{app:AnomalyDef}).}
\begin{equation}
\label{eq:EA}
 \mathcal{E}_{\mathrm{Weyl}}  \equiv \langle\tensor{T}{^\mu_\mu}\rangle_{\text{fin}} \;, \quad
 \mathcal{A}_{\mathrm{Weyl}} =  \frac{1}{\g}\left[g^{(d)}_{\mu\nu}\frac{\delta \Wct(d)}{\delta g_{\mu\nu}}\right]_{\text{fin}} \;,
\end{equation} 
where the subscript fin (div) indicates that only the finite (divergent) part is taken.
Combining with the first line of \eqref{eq:noCFT,A+E} one gets 
\begin{equation}
 \mathcal{A}_{\mathrm{Weyl}}    =g^{(4)}_{\mu\nu}\langle T^{\mu\nu}\rangle- \langle\tensor{T}{^\mu_\mu}\rangle_\fin + 
 \frac{g^{(4)}_{\mu\nu}}{\g}\frac{\delta \Wct(d)}{\delta g_{\mu\nu}} \;.
\end{equation}
If we further use  
\begin{equation}\label{eq:explicitbreakingct}
g^{(4)}_{\mu\nu}\frac{\delta \Wct(d)}{\delta g_{\mu\nu}}  = - \langle\tensor{T}{^\mu_\mu}\rangle_\Div
\end{equation}
which follows from the fact that the left-hand side vanishes in a classically Weyl invariant theory,
we then get
\begin{equation}
\label{eq:genacc}
 \mathcal{A}_{\mathrm{Weyl}}    =g^{(4)}_{\mu\nu}\langle T^{\mu\nu}\rangle- \langle\tensor{T}{^\mu_\mu}\rangle\; ,
\end{equation}
the formula \eqref{eq:AnomalyDef} which we promised to obtain by derivation. More details are provided in \APP\ref{app:AnomalyDef}. This formula is valid to all order in perturbation theory in dimensional regularisation and can be expected to be valid non-perturbatively.
Note that now the Weyl anomaly is not  the variation of a functional anymore, due to the explicit breaking in \eqref{eq:anomalyWren}. In that case, the Weyl anomaly does not generally respect the Wess-Zumino consistency conditions.

\section{Curved-space Anomaly  Constraints on the Weyl Anomaly}
\label{sec:gen}

The goal of  this section is to extend the results in  \cite{DUFF1977334} to theories  without classical Weyl invariance
and to consider the possibility of adding the  $P$- and $CP$-odd term  
\begin{equation}
R\tilde R \equiv \frac{1}{2}\epsilon^{\mu\nu\rho\sigma}\tensor{R}{^\alpha^\beta_\mu_\nu}\tensor{R}{_\alpha_\beta_\mu_\nu} \;,
\end{equation}
 to the counterterm action  given in \eqref{eq:Wctint}.\footnote{In another approach, 
the parity-odd term is investigated in a  conformal field theory \cite{Coriano:2023cvf} and it is found that it could 
appear in   3-point correlators with a marginal operator. We stress that this is a consistency constraint 
and not a proof of its existence.} 
  Unlike the other operators the extension of  
 $\RRt$ to $d$-dimension is ambiguous since the Levi-Civita tensor is intrinsically tied to $d=4$.
We may however proceed formally, without committing to a specific extension, as follows 
\begin{equation}
\Wctodd = \frac{1}{d-4}\int\dd^dx \,\g\,\Lagr_{\mathrm{odd}}(d)\;,
\end{equation}
with $\Lagr_{\mathrm{odd}}(4)  \propto \RRt$ as this is the only possible parity-odd term in $d=4$ 
(cf. Ref.\cite{Larue:2023tmu}).  Owing to its  topological nature one has
\begin{equation}\label{eq:deltaLodd4}
\frac{\delta}{\delta g^{(4)}_{\al\be}}\int\dd^4x\,\sqrt{g^{(4)}}\,\Lagr_{\mathrm{odd}}(4)=0\; .
\end{equation}
Note that \eqref{eq:deltaLodd4} does not imply that $\Lagr_{\mathrm{odd}}(d)$ is of $\ORD(d-4)$. For example, the extension of  the $P$-even topological Euler density to $d$ dimensions of  is discontinuous around $d=4$ \cite{Hennigar:2020lsl,Gurses:2020ofy}. However, we may parametrise the metric variation of $\Wctodd$ in terms of a continuous $(d-4)\times\mathcal{V}^{\al\be}$  and a discontinuous piece $\mathcal{U}^{\al\be}$
\begin{equation}\label{eq:deltaLodd}
\frac{1}{\g}\frac{\delta}{\delta g_{\al\be}}\int\dd^dx\,\g\,\Lagr_{\mathrm{odd}}(d)=(d-4) \mathcal{V}^{\al\be}(4)+\mathcal{U}^{\al\be}(d)+\mathcal{O}\left((d-4)^2\right)\; .
\end{equation}
The 2-tensors $\mathcal{V}^{\al\be}(4)$ and $\mathcal{U}^{\al\be}(d)$ are parity-odd, of mass dimension four, 
and one has
\begin{equation}
\mathcal{U}^{\al\be}(4)=0\;,\quad\quad \mathcal{U}^{\al\be}(4-2\eps)\neq\ORD(\eps)\;,
\end{equation}
owing to \eqref{eq:deltaLodd4}.
Using Bianchi identities, algebra, and intrinsically 4-dimensional identities \cite{Remiddi:2013joa,Chala:2021cgt,Chung:2022ees} 
one can show that the only possible operator (cf. \APP\ref{app:podd}) at $d=4$ is
\begin{equation}
\mathcal{V}^{\al\be}(4)=e\,g^{\al\be}R\tilde R\;,
\label{eq:Tctodd}
\end{equation}
where  $e$ is a constant to be determined.  For our purposes it is not necessary to 
specify the extension of the $\eps$-tensor to $d$ dimensions in parametrising 
  $\mathcal{U}^{\al\be}$ (this is similar to \cite{Elias:1982ea,Filoche:2022dxl} whereby one uses 
  free parameters in order to remain independent of a specific $\ga_5$-scheme) 
Using  Bianchi identities, which are $d$-independent, we may write
\begin{equation}
\mathcal{U}^{\mu\nu}(d)=e_1\, g^{\mu\nu}\RRt(d)+e_2\,P^{\mu\nu}(d)+e_3\,Q^{(\mu\nu)}(d)+e_4\,S^{(\mu\nu)}(d)\;,
\end{equation}
where round brackets denote symmetrisation $t^{(\mu\nu)}=\frac{1}{2}(t^{\mu\nu}+t^{\nu\mu})$ and
\begin{alignat}{4}
&g_{\al\be}R\tilde R(d)&\;=\;&g_{\al\be}\frac{1}{2}\epsilon^{\mu\nu\rho\sigma}R_{\gamma\delta\mu\nu}\tensor{R}{^\gamma^\delta_\rho_\sigma}\;,\quad\quad &P_{\alpha\beta}(d)&\;=\;&\epsilon^{\mu\nu\rho\sigma}R_{\alpha\lambda\mu\nu}\tensor{R}{_\beta^\lambda_\rho_\sigma}\nonumber\;, \\[0.1cm]
&Q_{\alpha\beta}(d)&\;=\;&\tensor{\epsilon}{_\alpha^\nu^\rho^\sigma}R_{\beta\nu\gamma\delta}\tensor{R}{_\rho_\sigma^\gamma^\delta}\;,\quad\quad &S_{\alpha\beta}(d)&\;=\;&\tensor{\epsilon}{_\alpha^\nu^\rho^\sigma}R_{\beta\lambda\rho\sigma}\tensor{R}{_\nu^\lambda}\nonumber\, ,
\end{alignat}
and any other $P$-odd symmetric 2-tensor is related to these by algebra and Bianchi identities (see \APP\ref{app:EMTconstraints}). 
At $d=4$, the Schouten identity reduces all these tensors to
\begin{equation}
P_{\mu\nu}(4)=Q_{\mu\nu}(4)=\frac{1}{2}\RRt(4)\;,  \quad S_{\mu\nu}(4)=0\;,
\end{equation}
and thus $\mathcal{U}^{\al\be}(4)=0$ implies the constraint
\begin{equation}\label{eq:Ualbe(d=4)}
2\,e_1+e_2+e_3=0\;.
\end{equation}
As expected, this defines a tensor that vanishes at $d=4$, but \emph{does not} scale as $\ORD(d-4)$.

Similarly, we can choose a basis of independent $P$-even operators, and then write a general ansatz for the metric variation of  $\Wct$ as follows
\begin{align}
\label{eq:Talphabeta}
\Tct^{\al\be}=\frac{1}{\g}\frac{\delta \Wct}{\delta g_{\al\be}}=&e\,g^{\al\be}R\tilde R+\frac{1}{d-4}\bigg\{e_1\,g^{\al\be}\RRt+e_2 \,P^{\al\be}+e_3\, Q^{(\al\be)}+e_4 \,S^{(\al\be)}\nonumber\\
& +a_1\,g^{\alpha\beta}R^2+a_2\,g^{\alpha\beta}R_{\mu\nu}R^{\mu\nu}+a_3\,g^{\alpha\beta}R_{\mu\nu\rho\sigma}R^{\mu\nu\rho\sigma}+a_4\,g^{\alpha\beta}\Box R\nonumber\\
&+b_1\,R R^{\alpha\beta}+b_2\, R^{\alpha\lambda}\tensor{R}{^\beta_\lambda}+b_3\,R_{\mu\nu}\tensor{R}{^\mu^\alpha^\nu^\beta}+b_4\,R^{\alpha\lambda\mu\nu}\tensor{R}{^\beta_\lambda_\mu_\nu}\nonumber\\
&+c_1\, D^{\alpha} D^{\beta}R+c_2\,\Box R^{\alpha\beta}\bigg\}\;,
\end{align}
We would like to add the following remarks:
\begin{itemize}
\item The  coefficients   $\{a_i,b_i,c_i\}$ are function of the parameters  in $\Wct$ and can depend on $d$,  such that the parity-even part $\Tct^{\al\be}$ can contain finite pieces despite the global $1/(d-4)$-factor.
\item The coefficients $\{e,e_i\}$ depend on the parameters in $\Wct$ and the scheme chosen for the $\eps$-tensor. For convenience, the $e_i$ are taken to be independent of $d$, since this would only amount to relabel $e$.
\item It is possible to write $\delta \Wct/\delta g_{\al\be}$ under this form because the counterterms $\Wct$ allowed by renormalisation are local polynomials. This is not the case of the effective action $\Weff$ which is non-local, e.g. \cite{DESER197645}.
\item The ansatz is valid for both theories with and without classical Weyl invariance.  
\item We only included tensors in $\Tct$, i.e covariant quantities. This does not amount to the absence of diffeo and Lorentz anomalies, but restricts them to their covariant form \cite{Bertlmann:1996xk}. 
\end{itemize}
We may trace $\Tct$ in \eqref{eq:Talphabeta}, using $g^{(d)}_{\mu\nu}g^{(d)\,\mu\nu}=d$, to 
obtain
\begin{align}\label{eq:traceT}
g^{(d)}_{\al\be}\Tct^{\al\be} &
 = (4\,e+e_1)\,R\tilde R +a_1\, R^2+a_2\,R_{\mu\nu}R^{\mu\nu}+a_3\,R_{\mu\nu\rho\sigma}R^{\mu\nu\rho\sigma}+a_4\,\Box R\nonumber\\
&+\frac{1}{d-4}\bigg\{(4\,a_1+b_1)R^2+(4\,a_2+b_2+b_3)R_{\mu\nu}R^{\mu\nu}\nonumber\\
&+(4\,a_3+b_4)R_{\mu\nu\rho\sigma}R^{\mu\nu\rho\sigma}+(4\,a_4+c_1+c_2)\Box R\bigg\}\;,
\end{align}
where we used
\begin{equation}
g^{(d)}_{\mu\nu}P^{\mu\nu}(d)=g^{(d)}_{\mu\nu}Q^{\mu\nu}(d)=2\RRt(d),\;\quad\quad g^{(d)}_{\mu\nu}S^{\mu\nu}(d)=0\;,
\end{equation}
upon using the Bianchi identities only. Incidentally, we see that there remains no $P$-odd term in the divergent part upon using \eqref{eq:Ualbe(d=4)}. In other words, the topological nature of the $P$-odd term in $\Wct$ (\EQ\eqref{eq:deltaLodd4}), forbids the presence of $P$-odd divergent terms in the trace of $\Tct$.

From \eqref{eq:traceT} we can compute the Weyl anomaly. 
That is
\begin{alignat}{3}
\label{eq:recall}
& \mathcal{A}_{\mathrm{Weyl}} &\;=\;&  [g^{(d)}_{\al\be}\Tct^{\al\be}]_{d=4}  \quad \quad & &    \text{classically Weyl invariant} \;, \nonumber \\[0.1cm]
& \mathcal{A}_{\mathrm{Weyl}} &\;=\;& [g^{(d)}_{\al\be}\Tct^{\al\be}]_{\text{fin},d=4}  \quad \quad   & & \text{classically Weyl non-invariant} \;, 
\end{alignat}
which follow from \EQs \eqref{eq:CFT,A,Wct}  and  \eqref{eq:EA} respectively.
In \SEC\ref{sec:diffeo} and \SEC\ref{sec:WZCC} the parameters $\{e,e_i\}$ and $\{a_i,b_i,c_i\}$ 
will be subjected to diffeo constraints and the Wess-Zumino consistency conditions.

\subsection{Constraints from the diffeomorphism anomaly}
\label{sec:diffeo}

The diffeo variation of the effective action is given by
\begin{align}
\label{eq:dWeff}
\delta^d_\xi \Weff=\int\dd^4 x\,\delta^d_\xi g_{\mu\nu}\frac{\delta \Weff}{\delta g_{\mu\nu}}=-\int\dd^4x\,\g\, \xi_\mu D_\nu\vev{T^{\mu\nu}}\;,
\end{align}
where $\delta^d_\xi$ is the (active) diffeo transformation, and we used $\delta^d_\xi g_{\mu\nu}=D_\mu \xi_\nu + D_\nu \xi_\mu$. 
Note that, \EQ\eqref{eq:dWeff} may be non-zero but remains finite 
and one-loop exact  (it is non-zero in theories 
in $d=2 + 4k$ with Weyl fermions for example \cite{Alvarez-Gaume:1983ihn}). 
 Since it is finite, it should not be altered by  $\Wct$  as there is no divergence to remove, 
 and therefore 
\begin{equation}
\delta^d_\xi \Wct=\int\dd^d x\,\delta^d_\xi g_{\mu\nu}\frac{\delta \Wct}{\delta g_{\mu\nu}}=-\int\dd^dx\,\g\, \xi_\mu D_\nu\Tct^{\mu\nu}=0\;,
\end{equation}
must hold. We thus compute the divergence of $\Tct$ and then enforce
\begin{equation}\label{eq:divT}
D_\nu\Tct^{\mu\nu}=0\;.
\end{equation}
The $P$-odd and -even pieces are independent of each other.    
 Enforcing  \eqref{eq:divT} on the $P$-even sector yields seven constraints
\begin{alignat}{6}
\label{eq:7}
&4\,a_1+b_1&\;=\;&0 \;, \quad 
& & 4\,a_2+b_2+b_3&\;=\;&0\;, \quad 
& & 4\,a_3+b_4&\;=\;&0 \nonumber\;, \\
& 4\,a_1+a_2+a_4&\;=\;&0\;, \quad 
& & 8\,a_3-b_2 &\;=\;&0\;, \quad 
& & 2\,a_2+8\,a_3+c_2 &\;=\;&0\nonumber\;, \\  
&4\,a_4+c_1+c_2 &\;=\;& -12\,a_1-4\,a_2-4\,a_3  \;. \!\! \!\!\!\!\!\!\!\!\!\!\!\!\!\!   \!\! \!\!\!\!\!\!\!\!\!\!\!\!\!\! \!\! \!\!\!\!\!\!\!\!\!\!\!\!\!\!  & & && && &   
\end{alignat}
Remarkably, the same procedure applied to the $P$-odd sector only admits the trivial solution
\begin{equation}
e=e_1=e_2=e_3=e_4=0\;,
\end{equation}
which implies the absence of $P$-odd operators in both the counterterms to the EMT, and the Weyl anomaly, for both classically Weyl invariant and non-invariant theories.
We stress that this result is independent of both, the underlying theory  and the $\eps$-tensor scheme. 
 In our previous 
work \cite{Larue:2023tmu} we have obtained the same result specifically for a theory with a  Weyl fermion only. Inserting these constraints in \eqref{eq:traceT} we obtain 
\begin{align}
\label{eq:traceT+diffeo}
g^{(d)}_{\al\be}\Tct^{\al\be} 
& = \frac{a_2+4\,a_3}{2}W^2 -\frac{a_2+2\,a_3}{2}E+\frac{3\,a_1+a_2+a_3}{3}R^2\;+ \\[0.1cm]
&\left( -(4\,a_1+a_2) -\frac{4}{d-4}(3\,a_1+a_2+a_3)\right) \Box R \;, \nonumber
\end{align}
an expression in terms of the three coefficients  $a_{1,2,3}$.  The $1/(d-4)$-term is addressed just below.

\subsection{Constraints from classical (non-)invariance }
\label{sec:WZCC}

As noted earlier anomalies are finite and thus the $1/(d-4)$-term in \eqref{eq:traceT+diffeo} ought to vanish. 

\subsubsection{Classical Weyl invariance  ($2$ anomaly coefficients)} 

The Weyl anomaly for a classically Weyl invariant theory is given by \EQ\eqref{eq:recall} 
and must respect the Wess-Zumino consistency condition.  In \EQ\eqref{eq:traceT+diffeo} it is only 
the $R^2$-term which does not respect the consistency conditions
 \cite{Bonora:1983ff,Bonora:1985cq,Cappelli:1988vw,Osborn1991WeylCC,Cappelli:1990yc}
 and therefore 
\begin{equation}
\label{eq:a123}
3\,a_1+a_2+a_3=0 \;,
\end{equation}
must hold which indeed  removes the divergent $\Box R$-term. 
 The result assumes the form 
\begin{equation}
\label{eq:A2}
\mathcal{A}_{\mathrm{Weyl}}  = a_1\, R^2+a_2\,R_{\mu\nu}R^{\mu\nu}- (3\, a_1 +a_2) \,R_{\mu\nu\rho\sigma}R^{\mu\nu\rho\sigma}- (4\, a_1+a_2) \,\Box R\;,
\end{equation}
which depends on the two  parameters $a_1$ and $a_2$  which are model-dependent.
This is the result obtained by Duff in \cite{DUFF1977334}, supplemented by 
the additional constraint that  $R\tilde R$ is absent. 

The question that poses itself is whether the 
results apply beyond LO. We would think that the answer is affirmative  to all orders in perturbation theory for  theories whose
non-anomalous symmetries are compatible with dimensional regularisation.\footnote{Supersymmetry might be such a counter-example since it is well-known to be incompatible with dimensional regularisation and also its supersymmetry-improved 
version of dimensional reduction \cite{Siegel:1980qs} (despite ongoing developments \cite{Stockinger:2005gx}).}
 Beyond LO,   $\Wct$ and the ansatz \eqref{eq:Talphabeta}  contains higher order poles in $d-4$. However, since the Weyl anomaly is finite, their contributions have to cancel 
 in analogy to the computation of an anomalous dimension of a parameter or an operator (e.g . \cite{Hathrell:1981zb}).

\subsubsection{Broken classical  Weyl symmetry ($3$ anomaly coefficients)}
In a theory that explicitly breaks Weyl invariance, the Wess-Zumino conditions do generally not apply. 
However,  since the Weyl anomaly  can be obtained as in   \EQ\eqref{eq:recall}, it is automatically given by the finite part and thus
\begin{equation}
\label{eq:A3}
\mathcal{A}_{\mathrm{Weyl}}  = a_1\, R^2+a_2\,R_{\mu\nu}R^{\mu\nu}+a_3\,R_{\mu\nu\rho\sigma}R^{\mu\nu\rho\sigma}- (4\,a_1+a_2) \,\Box R \;,
\end{equation}
the Weyl anomaly depends on the three parameters $a_{1,2,3}$ which are again model-dependent. 
Note that the $\Box R$-term is fixed in terms of others which we believe to be a new observation in the presence of explicit breaking.
With regards to the validity beyond LO the same remarks apply as in the previous section.

\subsection{Constraints from the Lorentz anomaly}
\label{sec:CL}

In the ansatz \eqref{eq:Talphabeta}, the assumption is made that the EMT is symmetric in its indices. To remain the most generic, we should assume that the theory may exhibit a Lorentz anomaly, which is the breaking of rotational symmetry at the quantum level and manifests itself in the antisymmetry of $\vev{T^{\mu\nu}}$. In practice, the antisymmetry can arise in the presence of  fermions since then the vierbein $\tensor{e}{^a_\nu}$ replace the metric, and the EMT is not automatically symmetric by metric-variation but follows from
\begin{equation}
\g\vev{T^{\mu\nu}}=\tensor{e}{^a^\nu}\frac{\delta \Weff}{\delta \tensor{e}{^a_\mu}}[\tensor{e}{^a_\mu}]\;,
\end{equation}
where $\tensor{e}{^a^\al}=g^{\al\be}\tensor{e}{^a_\be}$ with latin indices referring to the tangent frame. 
 The Lorentz anomaly is then defined by
 \begin{align}
 \label{eq:Ldef}
\int\dd^4x\,\g\,\al_{ab}\mathcal{A}^{ab}_{\mathrm{Lorentz}}=\delta^L_\al \Weff[\tensor{e}{^a_\mu}]=\int\dd^4x\,\g\,\al_{ab}\vev{T^{ab}}\;,
\end{align}
where   $\al_{ab}(x)=-\al_{ba}(x)$ is the Lorentz transformation  parameter, and 
$\delta^L_\al \tensor{e}{^a_\mu}=-\tensor{\al}{^a_b} \tensor{e}{^b_\mu}$ \cite{Bertlmann:1996xk} 
was made use of. Note that since the vierbein is not diffeo invariant, the diffeo anomaly becomes
\begin{equation}
\delta^d_\xi \Weff[\tensor{e}{^a_\nu}]=-\int\dd^4x\,\g\,\xi_\nu\left( D_\mu\vev{T^{\mu\nu}}-\tensor{\omega}{^\nu_a_b}\vev{T^{ab}}\right)\;,
\end{equation}
where $\omega_{\mu ab}=-\omega_{\mu ba} $ is the spin-connection.

The same argument as for the diffeo anomaly applies: since the Lorentz anomaly is finite, the counterterms must not contribute (i.e $\delta^L_\al \Wct=0$) yielding the Lorentz constraint
\begin{equation}
\Tct^{\al\be}=\Tct^{\be\al}\;.
\end{equation}
Enforcing the Lorentz constraint is equivalent to considering directly the metric variation as was done in this Section and \SEC\ref{sec:anomalydef}, since the vierbein variation can be split into its symmetric and antisymmetric parts $\delta \tensor{e}{^a_\nu}=\frac{1}{2}\delta g_{\mu\nu} e^{a \mu}-\tensor{\al}{^a_b}\tensor{e}{^b_\nu}$ \cite{Leutwyler:1985ar}. Besides, the diffeo constraint \eqref{eq:divT} is unchanged when the Lorentz constraint is verified.

\subsection{The non-removable $\Box R$-term}
\label{sec:boxR}

The constraints on the Weyl anomaly  in the case of a classically Weyl invariant theory imply that the coefficient of the $\Box R$ is fixed with respect to the coefficients of $R^2$, $R_{\mu\nu}R^{\mu\nu}$ and $R_{\mu\nu\rho\sigma}R^{\mu\nu\rho\sigma}$ \eqref{eq:A2} which has been known since a long 
time  \cite{DUFF1977334}. 
We believe that it is a new result that the same holds true for a theory which is not classically Weyl invariant \eqref{eq:A3}.  Further notice that the usual ambiguity in $\Box R$, due to the possibility of adding a local term 
\begin{equation}
\label{eq:dS}
 S_{R^2} =\int\dd^4x\,\g\,\al\,R^2\;, \quad    \tensor{T}{^\mu_\mu} \supset  \frac{2}{\g}   g_{\mu\nu}\frac{\de }{\delta g_{\mu\nu}}  S_{R^2} =-\frac{\al}{3}\Box R\; ,
\end{equation}
in the action, which shifts $\Box R$, is not present in the case at hand since this term cancels 
in the formula \eqref{eq:genacc}. This is the case since the term in \eqref{eq:dS} 
enters the explicit breaking $\mathcal{E}_{\text{Weyl}}$ and not the anomalous breaking in 
\eqref{eq:TEMT}. That is \eqref{eq:dS}
is intrinsically 4-dimensional and independent of the quantum fields, hence only affects the EMT at tree level.

The finding that the term \eqref{eq:dS} does not alter the anomalous part is related to  
 the possibility that a $\Box R$ flow theorem may exist \cite{Prochazka:2017pfa}. 
 However,  for the latter there is a problem when one considers the flow of QCD.  
 Pions cannot be coupled conformally, that is a quadratic term ${\cal L} \supset  \frac{1}{2}(D_\al \pi D^\al \pi  + \xi R \pi^2)$ 
 with $\xi = \frac{1}{6}$ (or any $\xi \neq 0$)  is not allowed by the shift symmetry for Goldstone bosons. 
 However $\xi \neq \frac{1}{6}$  renders the integral for the flow term infrared divergent.  
 In the purely anomalous part the non-conformal terms would drop out and suggest that 
 it might be worthwhile to consider whether one can formulate flow theorems  
 in terms of the anomalous part only.\footnote{Another alternative is that the theory has another Goldstone boson, the dilaton due to conformal symmetry breaking, in which case the pions can be coupled  conformally \cite{Zwicky:2023fay}. 
 Whether or not this is the case for low energy QCD is an open question and not generally believed to be the 
 case.} Effectively the infrared divergence is then shifted into the explicit breaking part which has no 
 relevance for the definition of the  (pure) Weyl anomaly. 
 
 Since we believe that the $\Box R$-anomaly  is calculable, the constraints on $\Box R$ in \eqref{eq:A2} and \eqref{eq:A3} 
 have to hold when computed with other regularisations provided they respect 
 the non-anomalous symmetries of the theory. 
 For example, the  $\zeta$-function regularisation employed in \cite{Fujikawa:1980rc,Fujikawa:2004cx} satisfies the constraint for a  spin-$1/2$ fermion and a spin-$1$ vector in the $(\frac{1}{2},0) \oplus (0,\frac{1}{2})$  and $(\frac{1}{2},\frac{1}{2})$ Lorentz representations respectively.   This is sometimes difficult to see when it is the total breaking $\mathcal{A}_{\mathrm{Weyl}} + \mathcal{E}_{\mathrm{Weyl}}$ that is given. That is the case for example for  the spin $0$ scalar  given in \cite{Birrell:1982ix}  (\TAB 1 in chapter 5). 
 However, since  $\xi = \frac{1}{6}$ removes the explicit breaking it is readily verified that in this case the constraint is satisfied. In \cite{Casarin:2018odz}, uniquely $\mathcal{A}_{\mathrm{Weyl}}$ is determined, and the constraint  is satisfied for any $\xi$.  Similarly, it was pointed in \cite{Casarin:2021fgd} that the heat kernel leads to a different definition  to \eqref{eq:AnomalyDef} in the presence of explicit breaking (the second term is replaced by $\vev{g^{(4)}_{\mu\nu}T^{\mu\nu}}$), and a different value for the $\Box R$. According to our derivation in \SEC\ref{sec:anomalyEX}, this implies that some explicit breaking is included in the heat kernel anomaly definition.
 Furthermore,  since ghosts and gauge fixing may affect  explicit breaking they require  careful assessment as well. 
 Hence, care must be taken when comparing the literature. Note also that the sign of $\Box R$ 
 is dependent on the sign convention of the metric.

\section{Curved-space Anomaly Including a Gauge Sector}
\label{sec:gauge}

In this section, we apply the same method in the presence of a gauge sector. 
This involves adding  gauge field terms in $\Wct$, or  the decomposition in  \eqref{eq:Talphabeta}. 

\subsection{Diffeomorphism anomaly with a gauge sector}
\label{sec:gaugediffeo}

In the presence of a background (abelian or non-abelian) gauge field $A_\mu$, the Lorentz and Weyl anomalies keep the same expression as in pure gravity 
since the gauge field is not affected by their transformation: 
 $\delta^W_\sig A_\mu=\delta^L_\al A_\mu=0$. The diffeo anomaly is subject to change because $A_\mu$ is not  diffeo invariant
\begin{equation}
\delta^d_\xi A^i_\mu=\xi^\nu\pa_\nu A^i_\mu+A^i_\nu\pa_\mu\xi^\nu \;.
\end{equation}
The gauge index $i$ is to be omitted for an  abelian gauge group.
Upon using $\delta^d_\xi \tensor{e}{^a_\mu}=\xi^\nu\pa_\nu \tensor{e}{^a_\mu}+\tensor{e}{^a_\nu}\pa_\mu\xi^\nu$ \cite{Bertlmann:1996xk,Fujikawa:2004cx}, 
one finds 
\begin{align}\label{eq:diffeoW+gauge}
& \delta^d_\xi \Weff[\tensor{e}{^a_\nu},A^i_\mu] =\int \dd^4x\,\left( \delta^d_\xi \tensor{e}{^a_\nu} \frac{\delta \Weff}{\delta \tensor{e}{^a_\nu}}+\delta^d_\xi A^i_\mu\frac{\delta \Weff}{\delta A^i_\mu} \right)  \\[0.1cm]
& =-\int\dd^4x\,\g\,\xi_\nu\Bigg\{ D_\mu\vev{T^{\mu\nu}}-\tensor{\omega}{^\nu_a_b}\vev{T^{ab}}
+A^{i,\nu}\left(D_\mu \frac{1}{\g}\frac{\delta \Weff}{\delta A_\mu}\right)^i+i\left(\tensor{F}{_\mu^\nu}\right)^i\frac{1}{\g}\frac{\delta \Weff}{\delta A^i_\mu}\Bigg\}\nonumber\;,
\end{align}
where $F_{\mu\nu}  =  \pa_\mu A_\nu -  \pa_\nu A_\mu +i [ A_\mu,A_\nu]$ is the gauge field strength. 
The third term can be written precisely as the gauge anomaly with $\theta^i \equiv \xi^\mu A^i_\mu$,   $\delta^g_\theta A^i_\mu=-(D_\mu\theta)^i$
\begin{equation}
\delta^g_\theta \Weff[\tensor{e}{^a_\nu},A^i_\mu]=\int\dd^dx\,\g\,\delta^g_\theta A^i_\mu \mathcal{H}^{i,\mu}=\int\dd^dx\,\g\,\theta^i\,\left(D_\mu \mathcal{H}^{\mu}\right)^i\;,
\end{equation}
using  
$\g\mathcal{H}^{i,\mu} \equiv \delta \Weff/\delta A^i_\mu$ as a shorthand.
Similar expressions arise for the counterterms $\Wct$ which have to be diffeo, Lorentz and gauge 
invariant as a consequence of the finiteness of these potential anomalies. 
Therefore, we may write the analogue of  \eqref{eq:diffeoW+gauge} in terms of $\Tct$ as
\begin{equation}
\label{eq:diffeoWct_wgauge}
D_\mu\Tct^{\mu\nu}-\tensor{\omega}{^\nu_a_b}\Tct^{ab}+A^{i,\nu}\left(D_\mu \mathcal{H}_{\text{ct}}^{\mu}\right)^i+i\left(\tensor{F}{_\mu^\nu}\right)^i\mathcal{H}_{\text{ct}}^{i,\mu}=0\;, 
\end{equation}
where $\g\mathcal{H}_{\text{ct}}^{i,\mu} \equiv \delta \Wct/\delta A^i_\mu$. The Lorentz and gauge constraints read
\begin{equation}
\Tct^{ab}=\Tct^{ba} \;,  \quad D_\mu \mathcal{H}_{\text{ct}}^{\mu}=0 \;,
\end{equation}
respectively such that when combined one finally gets 
\begin{equation}
\label{eq:DF}
 D_\mu\Tct^{\mu\nu}+i\left(\tensor{F}{_\mu^\nu}\right)^i\mathcal{H}_{\text{ct}}^{i,\mu}=0\;,
\end{equation}
the main equation of this section, representing the diffeo, the Lorentz and the gauge anomaly constraints  
 on the form of the Weyl anomaly.

\subsection{Constraints from the diffeomorphism anomaly on the gauge sector } 
\label{sec:CG}

In this section we apply the constraint \eqref{eq:DF}.
In the $P$-even case there exists a single counterterm 
\begin{equation}\label{eq:Wctgauge}
\Wct^{\mathrm{gauge}} =\frac{1}{d-4}\int\dd^dx\,\g\,f\,\tr\,\frac{1}{4}F^2\;,
\end{equation}
from which we deduce the metric variation
\begin{equation}
\Tct^{\al\be}=\frac{2}{\g}\frac{\delta \Wct^{\mathrm{gauge}}}{\delta g_{\al\be}}=\frac{1}{d-4}\,f\,\tr\,\left(\frac{1}{4}g^{\al\be}F^2-\tensor{F}{^\al^\lambda}\tensor{F}{^\be_\lambda}\right)\;,
\end{equation}
and gauge field variation
\begin{equation}
\mathcal{H}_{\text{ct}}^{i,\al}=\frac{1}{\g}\frac{\delta \Wct^{\mathrm{gauge}}}{\delta A^i_\al}=\frac{-i}{d-4}f\,\tr\,T^i\,D_\mu F^{\mu\al}\;,
\end{equation}
of the pure gauge part. 
The diffeo constraint \eqref{eq:DF} can be seen to be  satisfied (after using Bianchi identities) since \eqref{eq:Wctgauge} is indeed diffeo invariant.  The gauge field dependent part of the  Weyl anomaly is directly finite and reads
\begin{equation}
g^{(d)}_{\al\be}\Tct^{\al\be}=\frac{1}{d-4}f\,\tr\,\frac{d-4}{4}F^2=\frac{f}{4}\,\tr\,F^2\;.
\end{equation}
As is well-known, in the full   theory $f$ is proportional to the $\be$ function of the running gauge coupling.

Let us now consider the  more subtle case of $P$-odd operators. We proceed as for  pure gravity with 
the addition of gauge field dependent terms
\begin{equation}
\Wctodd^{\mathrm{gauge}}=\frac{1}{d-4}\int\dd^dx\,\g\,\mathcal{L}_{\mathrm{odd}}(d)\;.
\end{equation}
Since in $d=4$, the only $P$-odd gauge dependent operator of mass dimension 4 is the Pontryagin density $F\tilde F \equiv \frac{1}{2}\epsilon^{\mu\nu\rho\sig}F_{\mu\nu}F_{\rho\sig}$, we must have $\mathcal{L}_{\mathrm{odd}}(4)\propto\,\tr\,F\tilde F$.  Its topological nature implies
\begin{equation}
\frac{\delta}{\delta g_{\al\be}}\int\dd^4x\,\g\,\mathcal{L}_{\mathrm{odd}}(4) = \frac{\delta}{\delta A^i_{\al}}\int\dd^4x\,\g\,\mathcal{L}_{\mathrm{odd}}(4) =0 \;.
\end{equation}
With the same reasoning as for  pure gravity, we can express  metric and gauge variations of $\int\dd^dx\,\Lagr_{\mathrm{odd}}(d)$ in terms of a continuous and a discontinuous term
\begin{alignat}{2}
&\Tct^{\al\be}&\;=\;&\frac{1}{\g}\frac{\delta \Wctodd^{\mathrm{gauge}}}{\delta g_{\al\be}}=\mathcal{V}^{\al\be}(4)+\frac{1}{d-4}\mathcal{U}^{\al\be}(d)+\ORD(d-4) \;, \nonumber\\
&\mathcal{H}_{\text{ct}}^{i,\al}&\;=\;&\frac{1}{\g}\frac{\delta \Wctodd^{\mathrm{gauge}}}{\delta A^i_\al}=\mathcal{M}^{i,\al}(4)+\frac{1}{d-4}\mathcal{N}^{i,\al}(d)+\ORD(d-4)\;,
\end{alignat}
with
\begin{equation}
\mathcal{U}^{\al\be}(4) = 0\;,\quad\quad \mathcal{U}^{\al\be}(4-2\eps) \neq \ORD(\eps) \;,
\end{equation}
and likewise for ${\cal N}$. 
Using the intrinsically 4-dimensional Schouten identity, one can show that the only possible $P$-odd gauge covariant 2-tensor of mass dimension four is
\begin{equation}
\mathcal{V}^{\al\be}(4)=h\,g^{\al\be}\tr\,F\tilde F\;,
\end{equation} 
which is the analogue of \eqref{eq:Tctodd}.
Without assuming a specific scheme for the  $\eps$-tensor, we may write without loss of generality 
\begin{equation}
\mathcal{U}^{\al\be}(d)=h_1\,g^{\al\be}\tr\,F\tilde F+h_2\,\tr\,\tensor{F}{^(^\al_\lambda}\tensor{\tilde F}{^\be^)^\lambda}\;,
\end{equation}
using $d$-independent identities.
Note that the trace has to be ignored if the gauge group is abelian.
We only consider symmetric tensors such that the Lorentz constraint (cf. \SEC\ref{sec:CL}) is directly enforced.

At $d=4$, using the Schouten identity we obtain ($\mathcal{U}^{\al\be}(4)=0)$ 
\begin{equation}\label{eq:gaugeUd=4}
 (4\,h_1+h_2)g^{\al\be}\tr\,F\tilde F = 0  \quad \Rightarrow  \quad 4\,h_1+h_2=0\;.
\end{equation}
In principle, one should include mixed gravity-gauge 2-tensors in $\mathcal{V}^{\al\be}(4)$ and $\mathcal{U}^{\al\be}(d)$. They are however independent from the pure gauge operators, and vanish under the metric contraction, hence cannot contribute to the Weyl anomaly (see \APP\ref{app:Poddgauge}).
 
Turning to $\mathcal{H}_{\text{ct}}$, the second term in  our constraint equation \eqref{eq:DF}, we first notice that 
these mixed gravity-gauge terms can be ignored for the same reason. Secondly, the only $P$-odd $1$-tensor of mass dimension $3$ we could be tempted to write is
\begin{equation}
\mathcal{N}^{i,\al}(d)=h'\,\epsilon^{\al\nu\rho\sig}\tr\,T^i\,D_\nu F_{\rho\sig}=0\;,
\end{equation}
which vanishes by virtue of the Bianchi identity. As a result we also have $\mathcal{M}^{i,\al}=0$.

The  gauge dependent piece of the Weyl anomaly using this parametrisation is
\begin{equation}
g^{(d)}_{\al\be}\Tct^{\al\be}=(4\,h+h_1)\tr\,F\tilde F\;,
\end{equation}
where like in the pure gravity case,  the divergent part is zero by   $4\,h_1+h_2=0$ \eqref{eq:gaugeUd=4}.
Finally, the diffeo constraint \eqref{eq:DF} on the pure gauge terms reads
\begin{equation} 
D_\mu\Tct^{\mu\nu}+i\left(\tensor{F}{_\mu^\nu}\right)^i \mathcal{H}_{\text{ct}}^{i,\mu}=0\;,
\end{equation}
and leads to a system that only admits the solution
\begin{equation}
h=h_1=h_2=0\;.
\end{equation}
We thus conclude on the absence of $P$-odd gauge dependent terms in the counterterms and the Weyl anomaly, for both non-abelian and abelian gauge groups.

\section{Explicit Gravity-gauge  Anomalies for a  Weyl fermion} 
\label{sec:mixed}

In this section, we compute Weyl, diffeo and Lorentz anomalies for a $(\frac{1}{2},0)$ Weyl fermion 
including a gauge sector. Note that a Weyl fermion is not the same as a Dirac fermion coupled chirally to gauge or gravitational fields \cite{Armillis:2010pa} (where parity odd terms are absent) as emphasised in  \cite{Larue:2023tmu}.
Since these anomalies involve gauge and gravitational fields,  they are sometimes 
referred to as mixed gravity-gauge anomalies in the literature.
We follow the notation in our previous work  \cite{Larue:2023tmu}.
The action of the Weyl fermion is given by 
\begin{equation}\label{eq:Weylfermion}
S=\int\dd^4x\,\tilde{\bar\psi}_L \DM \tilde\psi_L\;,
\end{equation}
where $\tilde\psi_L=\sqrt{e}\psi_L$   is the two-component Weyl fermion 
appropriate for avoiding  spurious diffeo anomalies  in the path integral \cite{Fujikawa:1980rc,Toms:1986sh,Fujikawa:2004cx}. 
Here, the variable $e= \det\tensor{e}{^a_\mu}=\g$ is the vierbein determinant.
The Dirac-Weyl operator is defined by 
\begin{equation}\label{eq:DW2}
\DM \equiv i\bar\sigma^\mu D_\mu  \;, \quad 
 D_\mu \tilde\psi_L \equiv  \left(\pa_\mu+\omega^L_\mu-\frac{\pa_\mu\sqrt{e}}{\sqrt{e}}+i A_\mu\right)\tilde\psi_L \;,
\end{equation}
where  $A_\mu$ is the gauge field (abelian or non-abelian),  $\omega^L_\mu=\frac{1}{8}\omega_{\mu ab}(\sigma^a\bar\sigma^b-\sigma^b\bar\sigma^a)$  the spin connection, 
$\bar\sig^\mu = \tensor{e}{^\mu_a}\bar \sig^a$ and  $\sig(\bar \sig)^a = (1, \pm \vec{\sig})$ where $\vec{\sig}$ are the Pauli matrices.   The effective action, with Euclidian signature,  assumes the form
\begin{equation}
\Weff[\tensor{e}{^a_\mu},A_\mu]=-\log\,\int\DP\tilde{\bar\psi}_L\DP\tilde\psi_L e^{-S}=-\log\det \DM\;.
\end{equation}
For Dirac fermions  the functional determinant of an operator is defined by expressing it as the product of its eigenvalues. In our case,  $\DM$ maps left-handed onto right-handed Weyl fermions and 
hence the eigenvalue equation $\DM \phi=\lambda\phi$ is meaningless \cite{Alvarez-Gaume:1983ihn}.
Even though $\Weff$ is ill-defined, its variation
\begin{equation}
\label{eq:Weff}
\delta \Weff = -\Tr \,\delta \DM \DM^{-1}\;,
\end{equation}
is well-defined, since  $\delta \DM \DM^{-1}$ maps fermions of the same chirality into each other 
\cite{Leutwyler:1984de,Leutwyler:1985ar}.
This quantity can be regularised by introducing a smooth function $f$, with  $f(0)=1$ and $\forall\, n\geq0$, $\lim_{x\to\infty}x^n f^{(n)}(x)=0$, and making use of the formal expression \cite{Leutwyler:1985ar}
\begin{equation}
\label{eq:Dm}
\left.\DM^{-1}\right|_\Lambda=\DM^\dagger \int_{1/\Lambda^2}^\infty \dd t\,  f'(-t\,\DM\DM^\dagger)\underset{\Lambda\to\infty}{\longrightarrow}\DM^{-1}\;.
\end{equation}
This is well-defined since $\DM\DM^\dagger$ is a positive operator provided that the zero modes are excluded (i.e perturbative set-up). For the practical computation we will use $f(x)=1/(1+x)$, suitable 
for the Covariant Derivative Expansion (CDE)   technique for curved space (e.g. \cite{Larue:2023uyv}). 
Finally, from \eqref{eq:Weff} and \eqref{eq:Dm} one infers 
the regularised effective action variation
\begin{equation}\label{eq:deltaWreg}
\delta \Weff = -\lim_{\Lambda\to\infty}\Tr \,\delta \DM \DM^{-1}f\left(-\frac{\DM\DM^\dagger}{\Lambda^2}\right)\,,
\end{equation}
which will be the basis for the computations in the remaining part of this 
section.

\subsection{Weyl, diffeomorphism, Lorentz and gauge anomalies}

From  the Weyl-, diffeo- and Lorentz-variations of the background fields $\tensor{e}{^a_\mu}$ and $A_\mu$, we obtain the Dirac-Weyl operator variations \cite{Larue:2023tmu}
\begin{alignat}{2}
\label{eq:old}
& \delta^W_\sig \DM &\;=\;& -\sig\DM-\frac{1}{2}[\DM,\sig] \nonumber \\
& \delta^d_\xi \DM &\;=\;& -[\DM,\xi^\mu\nabla_\mu]-\frac{1}{2}[\DM,(D_\mu \xi^\mu)]  \nonumber \\
&\delta^L_\al\DM &\;=\;& \frac{1}{2}[\DM,\al_{ab}\mu^{ab}]+\frac{1}{2}\al_{ab}(\mu^{ab}-\lambda^{ab})\DM\;,
\end{alignat}
where  $\lambda^{ab} \equiv \frac{1}{4}(\bar\sigma^a\sigma^b-\bar\sigma^b\sigma^a)$,  $\mu^{ab} \equiv \frac{1}{4}(\sigma^a\bar\sigma^b-\sigma^b\bar\sigma^a)$ and $\nabla_\mu$ is the diffeo-covariant 
derivative.\footnote{$\nabla$ does neither include $A_\mu$ nor $\omega_\mu$, it only contracts indices. For example we have $D_\mu v_\nu=\nabla_\mu v_\nu +[\omega_\mu,v_\nu]+[i A_\mu,v_\nu]$ with $\nabla_\mu v_\nu=\pa_\mu v_\nu-\Gamma^\rho_{\mu\nu}v_\rho$. In particular we have $\nabla_\mu\tilde\psi=\pa_\mu\tilde\psi-\frac{\pa_\mu\sqrt{e}}{\sqrt{e}}\tilde\psi=\sqrt{e}\pa_\mu\psi$.}  The gauge variation, not treated in \cite{Larue:2023tmu}, reads 
\begin{equation}
\label{eq:gauge}
\delta^g_\theta \DM = -[\DM,\theta]\; .
\end{equation}
When inserting these expressions in \eqref{eq:deltaWreg} it turns out that one can recast the equations 
in terms of the Dirac operator \cite{Larue:2023tmu}
\begin{equation}
\sD  \equiv \gamma^\mu \left(\pa_\mu+\omega_\mu-\frac{\pa_\mu\sqrt{e}}{\sqrt{e}}+i A_\mu\right)\;,
\end{equation}
where $\omega_\mu \equiv \frac{1}{8}\omega_{\mu ab}[\gamma^a,\gamma^b]$.
We thus obtain 
\begin{alignat}{2}\label{eq:anomalies}
& \delta^W_\sigma \Weff &\;=\;& \frac{1}{2}\lim_{\Lambda\to\infty}\Tr\,\sig \frac{\Lambda^2}{\Lambda^2+ \sD^2} 
\nonumber \;, \\[0.1cm]
& \delta^{d}_\xi \Weff &\;=\;& -\lim_{\Lambda\to\infty}\Tr\, \gamma_5\left(\xi^\mu \nabla_\mu+\frac{1}{2}(D_\mu\xi^\mu)\right) \frac{\Lambda^2}{\Lambda^2+ \sD^2} \nonumber \;, \\[0.1cm]
& \delta^L_\al \Weff &\;=\;& \lim_{\Lambda\to\infty}\Tr\,\frac{1}{8}\alpha_{ab}[\gamma^a,\gamma^b]\gamma_5 \frac{\Lambda^2}{\Lambda^2+ \sD^2}\nonumber \;, \\[0.1cm]
& \delta^g_\theta \Weff &\;=\;&  -\lim_{\Lambda\to\infty}\Tr\,\theta\gamma_5 \frac{\Lambda^2}{\Lambda^2+ \sD^2}\;.
\end{alignat}
 Note that by writing $\nabla_\mu=D_\mu-\omega_\mu-i A_\mu$ in 
 the second line of \eqref{eq:anomalies}, the diffeo anomaly can be recast 
\begin{equation}
 \delta^{d}_\xi \Weff = -\lim_{\Lambda\to\infty}\Tr\, \gamma_5\left(\xi^\mu D_\mu+\frac{1}{2}(D_\mu\xi^\mu)\right) \frac{\Lambda^2}{\Lambda^2+ \sD^2}-\delta^g_{\xi^\mu iA_\mu}\Weff+\delta^L_{\xi^\mu \omega_{\mu ab}}\Weff\; ,
\end{equation}
in terms of the covariant derivative and  gauge and Lorentz anomalies (with   
  parameters $\theta=\xi^\mu iA_\mu$  and 
$\al_{ab}=\xi^\mu \omega_{\mu ab}$ respectively).
The appearance of the gauge anomaly, as noticed earlier, is due the  gauge field being diffeo-variant.  
We Wick rotate back to Minkowski signature for the following results.
For the Weyl anomaly we obtain using the CDE
\begin{equation*}
 \delta^W_\sigma \Weff \;=\; \frac{1}{32\pi^2}\int\dd^4x\,\sqrt{-g}\,\sig\,\bigg(\frac{1}{72}R^2-\frac{1}{45}R_{\mu\nu}R^{\mu\nu}-\frac{7}{360}R_{\mu\nu\rho\sigma}R^{\mu\nu\rho\sigma} 
-\frac{1}{30}\Box R -\frac{2}{3}\,\tr\,F^2\bigg) \;,
\end{equation*}
where the sign in $F^2$ differs from some of the literature since $F$ is defined in \SEC\ref{sec:gaugediffeo} with an extra factor of $i$ with respect to the (standard) convention.
This extends the results from \cite{Larue:2023tmu} to include a gauge sector.  
In analogy to the absence of the $\RRt$-term we  find by explicit computation that the $F\tilde F$-term
is absent, in agreement  with \SEC\ref{sec:CG}. 
The other  anomalies in  \eqref{eq:anomalies} evaluate to  
\begin{alignat}{2}
\label{eq:anomaliesB}
& \delta^{d}_\xi \Weff &\;=\;&-\delta^g_{\xi^\mu iA_\mu}\Weff+\delta^L_{\xi^\mu \omega_{\mu ab}}\Weff+\frac{1}{32\pi^2}\int\dd^4x\,\sqrt{-g}\,(D^\nu\xi^\mu)\epsilon^{\al\be\rho\sig}R_{\mu\nu\rho\sig}\,\tr\,F_{\al\be}\nonumber \;, \\[0.1cm]
& \delta^L_\al \Weff &\;=\;&\frac{-1}{32\pi^2}\int\dd^4x\,\sqrt{-g}\,\al_{\mu\nu}\epsilon^{\mu\nu\rho\sig}\left(\frac{1}{6}\,\tr\,\Box F_{\rho\sig}+\frac{1}{12}\tensor{R}{^\al^\be_\rho_\sig}\,\tr\,F_{\al\be}-\frac{1}{12}R\,\tr\,F_{\rho\sig}\right)\nonumber \;, \\[0.1cm]
& \delta^g_\theta \Weff &\;=\;& \frac{-i}{16\pi^2}\int\dd^4x\,\sqrt{-g}\,\tr\,\theta\,\left(\frac{1}{24}R\tilde R-F\tilde F \right)\;,
\end{alignat} 
where $\al_{\mu\nu}=\al_{ab}\,\tensor{e}{^a_\mu}\tensor{e}{^b_\nu}$. Traces run over gauge 
indices (and is to be ignored if the gauge group is abelian), and the same remark with respect to the sign of $F\tilde F$ applies as above.\footnote{Note that we have obtained the covariant anomalies since our method 
uses  $\DM\DM^\dagger$ in \eqref{eq:deltaWreg} which can be expressed 
in terms of the field strength and the Riemann tensor.
 Had we chosen a regularisation that involves $\DM^2$ instead, we would obtain the consistent form of the anomalies only \cite{Bertlmann:1996xk,Filoche:2022dxl}.}   
 Note that a chiral Dirac fermion coupling to $A$ as $S_D=\int\tilde{\bar\psi}i\left(\slashed\nabla+\slashed\omega +i\slashed A P_L\right)\tilde\psi$ has the same mixed anomalies (and only the mixed ones) as the Weyl fermion  \eqref{eq:Weylfermion}. Indeed $S_D$ is the sum of \eqref{eq:Weylfermion}, and of a right-handed Weyl fermion not coupling to $A$ which produces no additional mixed anomalies. 
The Lorentz anomaly  in \eqref{eq:anomaliesB} agrees with  \cite{Nieh:1984vx,Caneschi:1985an,Yajima:1986bi}. The gauge anomaly is standard, the $R\tilde R$-term corresponds to the mixed-axial gravitational anomaly \cite{Bertlmann:1996xk,Fujikawa:2004cx} and vanishes if the gauge group is non-abelian. 
To the best of our knowledge, this is the first time that the diffeo anomaly of a Weyl fermion is explicitly computed and  shown to bear mixed terms (the same applies to a chiral Dirac fermion as mentioned earlier).
In \cite{Caneschi:1985an}  the divergence of the EMT was considered which is only a part of the diffeo anomaly \eqref{eq:diffeoW+gauge}.\footnote{The mixed anomalies in the different currents are related by 
local counterterms \cite{Caneschi:1985an}. However as pointed out in \cite{LEUTWYLER198565,Leutwyler:1985ar}, the local counterterm that relates  Lorentz and diffeo anomalies \cite{Alvarez-Gaume:1983ict,Bardeen:1984pm} is non-polynomial, hence not permitted by the rules of renormalisation.}
 
\subsection{Phenomenological considerations}

Note that whereas the results for the Weyl anomaly are physical, the other anomalies \eqref{eq:anomaliesB}  must cancel in a ultraviolet complete theory.  
Another difference is that the anomalies in \eqref{eq:anomaliesB} are one-loop (or LO) exact 
where the Weyl anomaly  has in general  contributions to all orders and beyond.  
 The main points for abelian and non-abelian cases are:
\begin{itemize}
 \item [a)] Abelian case:  the Lorentz and diffeo anomaly cancellation requires 
$ \sum_i  Q_i = 0$ where $Q_i$ are the $U(1)$-charges of  the Weyl fermions 
(or the  Dirac fermions coupled 
to the $U(1)$ via a chiral projector).
If we had right-handed fermions in addition then the condition would read: $ \sum_{i}  Q_{L_i} - Q_{R_i} = 0$. The diffeo and the Lorentz anomaly cancellation thus gives the same contraints as the gauge anomaly cancellation.
\item [b)]   Non-abelian case: the Lorentz anomaly vanishes and the gauge anomaly is equivalent 
to the standard one. However, the diffeo anomaly does not vanish due to the $\delta^g_{\xi^\mu A^i_\mu}W$-term. For a gauge group $SU(n)$ this leads to the condition 
$\tr [T^a \{ T^b,T^c\}] \propto d^{abc} =0$, often 
referred to as $SU(n)^{\otimes 3}$-anomaly condition. Hence the diffeo anomaly cancellation condition is the same as the gauge anomaly one.
 \end{itemize}

As emphasised above there are effectively no new gauge constraints. 
However, in the case where one has a   vector field  associated to a global chiral symmetry  there 
are interesting phenomenological  consequences. Let us consider 
  a global chiral vector field. Whereas the gauge anomaly $\delta^g_\theta W$ \eqref{eq:anomaliesB}
  is not relevant, the Lorentz and diffeo anomalies are and thus the constraints found in a) and b) do apply. 
 Should the chiral field theory not satisfy these constraints this implies that the theory cannot be 
 ultraviolet  complete. Examples of where such models have been considered include 
 \cite{Hambye:2008bq,Chaffey:2019fec,Dror:2017ehi,Dror:2017nsg,Dror:2018wfl}. 
 However, these models are still meaningful as effective  theories since they 
 can arise as  low energy theories of a Higgs mechanism where the anomaly-cancelling fermions 
 have been integrated out \cite{Ball:1988xg,Preskill:1990fr}. {The only deficiency of such theories 
 are that they are not renormalisable and have a maximum energy up to which they can be used.} 
 In these references only gauge anomalies were considered but 
 it would seem that the very same reasoning does equally apply to the case of Lorentz and diffeo anomalies.

\section{Summary and Conclusion} 
\label{sec:conc}

In this work we have revisited the constraints on the EMT arising from the finiteness of the 
diffeomorphism, the Lorentz and the gauge anomaly. Our work extends Ref.~\cite{DUFF1977334} to include parity-odd terms and explicit symmetry breaking. 
 We  derive the formula for the pure Weyl anomaly  \eqref{eq:genacc}, widely used in the literature,  within dimensional regularisation.
 The most general decomposition of the 
 EMT counterterms \eqref{eq:Talphabeta} consists of ten parity-even and one parity-odd  term. After 
 applying the diffeomorphism anomaly and Wess-Zumino consistency constraints it is found that the parity-odd 
 term vanishes and that the parity-even terms reduce to three (two) unknowns in the case of (no-)explicit 
 symmetry breaking.  In particular, this implies the absence of the $\RRt$-term from the Weyl anomaly 
 in a model-independent way.
 The $\Box R$-anomaly  is found to be  finite, unambiguous and determined 
 through the Weyl and Euler coefficients. That this still holds in the case of explicit Weyl breaking is a new result
 (\SEC\ref{sec:boxR}).  It was argued that 
  the constraints should extend to all orders 
  in  theories whose non-anomalous symmetries are respected by dimensional regularisation.
  In \SEC\ref{sec:gauge} the method
 was extended to include a gauge sector, ruling out the $F\tilde F$-term.
 Mixed gravity-gauge anomalies  
 for Weyl fermions were computed to leading order (\SEC\ref{sec:mixed}).  We believe that our results for the diffeomorphism anomaly are more 
 complete than in  the literature. We deduce phenomenological constraints on UV-completeness  of theories that include
 vector bosons associated with global chiral  symmetries.

\section*{Acknowledgments} 
The authors are grateful to Latham Boyle, Franz Herzog and Neil Turok
 for helpful discussions, and we acknowledge valuable comments by the referee. The work of RL and JQ is supported by the project AFFIRM of the Programme National GRAM of CNRS/INSU with INP and IN2P3 co-funded by CNES and by the project EFFORT supported by the programme IRGA from UGA. JQ and RZ acknowledge the  support of CERN associateships. The work of RZ is supported by the  STFC Consolidated Grant, ST/P0000630/1. 
Many manipulations were carried out with the help of Mathematica and the package xAct \cite{xAct}.

\appendix

\section{Anomaly Definition}
\label{app:AnomalyDef}

The Weyl variation of $\Wren(d)$ is defined by \eqref{eq:anomalyWren}  with explicit symmetry breaking, and   \eqref{eq:anomalyWrenCFT} without. Either way, if we take the limit before the Weyl variation we obtain
\begin{align}
\delta^W_\sig \lim_{d\to4} \Wren(d)=\int\dd^4x\,\sig g^{(4)}_{\mu\nu}\frac{\delta}{\delta g_{\mu\nu}}\left[\Wren(d)\right]_{d=4}=\int\dd^4x\,\sig \left[ g^{(4)}_{\mu\nu}\frac{\delta \Weff(d)}{\delta g_{\mu\nu}} + g^{(4)}_{\mu\nu}\frac{\delta \Wct(d)}{\delta g_{\mu\nu}}\right]_{d=4}\nonumber\;,
\end{align}
where we used the fact that $\Wren(4)$ is finite to commute the metric variation and the metric contraction inside the limit. Alternatively, the Weyl anomaly can be written by taking the limit after the Weyl variation
\begin{equation*}
\lim_{d\to4} \delta^W_\sig \Wren(d)=\int\dd^4x\,\sig \left[g^{(d)}_{\mu\nu}\frac{\delta \Wren(d)}{\delta g_{\mu\nu}}\right]_{d=4}=\int\dd^4x\,\sig \left[ g^{(d)}_{\mu\nu}\frac{\delta W(d)}{\delta g_{\mu\nu}} + g^{(d)}_{\mu\nu}\frac{\delta \Wct(d)}{\delta g_{\mu\nu}}\right]_{d=4}\;.
\end{equation*}
In the following we prove some identities on the counterterms that are used in the derivation of \EQs \eqref{eq:anomalyWCFT}, \eqref{eq:CFT,A,Wct}, \eqref{eq:EA} and \eqref{eq:genacc}.

\subsection{Classical Weyl invariance } 

The most generic form of the counterterms in a classically Weyl invariant theory reads \cite{DUFF1977334}
\begin{equation}
\Wct(d)=\frac{1}{d-4}\int\dd^d x\, \g \,\left(a\,E(d) + c\, C(d) + f\,\tr\,F(d)^2\right)\;,
\end{equation}
where we included only $P$-even terms. We omit $P$-odd operator since we have showed that they must be absent at the level of the metric variation of $\Wct$.  
One can show that \cite{DUFF1977334} 
\begin{alignat}{2}
& \frac{2}{\g}g^{(d)}_{\mu\nu}\frac{\delta}{\delta g^{(d)}_{\mu\nu}}\int\dd^dx\,\g\,C&\;=\;&(d-4)\left(C+\frac{2}{3}\Box R\right) \;,\nonumber\\
&\frac{2}{\g}g^{(d)}_{\mu\nu}\frac{\delta}{\delta g^{(d)}_{\mu\nu}}\int\dd^dx\,\g\,E&\;=\;&(d-4)E \;, \nonumber\\
&\frac{2}{\g}g^{(d)}_{\mu\nu}\frac{\delta}{\delta g^{(d)}_{\mu\nu}}\int\dd^dx\,\g\,F^2&\;=\;&(d-4)F^2 \label{app:Wctvariation1}\;.
\end{alignat}
Whereas contracting with the metric in $d=4$ one obtains  \cite{DUFF1977334}
\begin{alignat}{2}
& \frac{2}{\g}g^{(4)}_{\mu\nu}\frac{\delta}{\delta g^{(d)}_{\mu\nu}}\int\dd^dx\,\g\,C&\;=\;&0 \;, \nonumber\\
&\frac{2}{\g}g^{(4)}_{\mu\nu}\frac{\delta}{\delta g^{(d)}_{\mu\nu}}\int\dd^dx\,\g\,E&\;=\;&0  \;,\nonumber\\
&\frac{2}{\g}g^{(4)}_{\mu\nu}\frac{\delta}{\delta g^{(d)}_{\mu\nu}}\int\dd^dx\,\g\,F^2&\;=\;&0\label{app:Wctvariation2}\;,
\end{alignat}
which then can be used to deduce that 
\begin{equation}
\left[ g^{(4)}_{\mu\nu}\frac{\delta \Wct}{\delta g_{\mu\nu}} \right]_{d=4} =0\;,
\end{equation}
holds in a classically Weyl invariant theory. For example for the Euler density we find
\begin{align}
\frac{\delta}{\delta g^{(d)}_{\mu\nu}}\int\dd^dx\,\g\,E=&8\,R^{\mu\al}\tensor{R}{^\nu_\al}-4\,R^{\mu\nu}R+8\,R^{\al\be}\tensor{R}{^\mu_\al^\nu_\be}-4\,R^{\mu\al\be\de}\tensor{R}{^\nu_\al_\be_\de}\nonumber\\
&+g^{\mu\nu}(R^2-4\,R_{\al\be}R^{\al\be}+R_{\al\be\ga\de}R^{\al\be\ga\de})\;,
\end{align}
where all the tensors are in $d$. We then obtain the above result for $E$ upon using\\ $g^{(d),\mu\nu}g^{(d)}_{\mu\nu}=d$ and $g^{(4),\mu\nu}g^{(d)}_{\mu\nu}=4$.

\subsection{Classical  Weyl non-invariance} 

The Weyl variation after and before taking the $d\to4$ limit read
\begin{alignat}{2}
&\delta^W_\sig \lim_{d\to4} \Wren (d) &\;=\;&\int\dd^4x\,\sig \left[\g g^{(4)}_{\mu\nu}\langle T^{\mu\nu}\rangle+g^{(4)}_{\mu\nu}\frac{\delta \Wct (d)}{\delta g_{\mu\nu}}\right]_{d=4} \nonumber \\[0.1cm]
&\lim_{d\to4}\delta^W_\sigma \Wren(d) &\;=\;&\int\dd^4x\,\sig \left[\g\langle\tensor{T}{^\mu_\mu}\rangle+g^{(d)}_{\mu\nu}\frac{\delta \Wct(d)}{\delta g_{\mu\nu}}\right]_{d=4}\label{app:deltaWren2}\;,
\end{alignat}
where in each line, the sum is finite but each term is separately divergent. Let us separate
\begin{alignat}{2}
&\langle\tensor{T}{^\mu_\mu}\rangle&\;=\;&\langle\tensor{T}{^\mu_\mu}\rangle_{\text{div}}+\langle\tensor{T}{^\mu_\mu}\rangle_{\text{fin}}\label{eqapp:Tmumudivfin} \nonumber  \\[0.1cm]
& g^{(d)}_{\mu\nu}\frac{\delta \Wct(d)}{\delta g_{\mu\nu}} &\;=\;& \left[g^{(d)}_{\mu\nu}\frac{\delta \Wct(d)}{\delta g_{\mu\nu}}\right]_{\text{div}}+\left[g^{(d)}_{\mu\nu}\frac{\delta \Wct(d)}{\delta g_{\mu\nu}}\right]_{\text{fin}}\;,
\end{alignat}
into their finite and divergent pieces.
We then identify the explicit breaking with
\begin{equation}\label{appeq:EWct}
 \mathcal{E}_{\mathrm{Weyl}}\equiv\vev{\tensor{T}{^\mu_\mu}}_{\text{fin}}\;,
\end{equation}
from which we can deduce the expressions for the anomaly in terms of $\Wct$
\begin{equation}\label{appeq:AWct}
 \mathcal{A}_{\mathrm{Weyl}} =  \frac{1}{\g}\left[g^{(d)}_{\mu\nu}\frac{\delta \Wct(d)}{\delta g_{\mu\nu}}\right]_{\text{fin}} \;,
\end{equation}
or equivalently in terms of the effective action $\Weff$
\begin{equation}\label{eqapp:AETM}
\mathcal{A}_{\mathrm{Weyl}}    =g^{(4)}_{\mu\nu}\langle T^{\mu\nu}\rangle- \langle\tensor{T}{^\mu_\mu}\rangle\; .
\end{equation}

\paragraph{Finite piece in $\Wct$}
Let us first provide more details about the comment in footnote \ref{foot:finiteterminWct}. One may argue regarding \eqref{appeq:AWct} that introducing a finite piece in $\Wct$ will alter the expression of the anomaly. In fact \eqref{appeq:AWct} and \eqref{appeq:EWct} have to be amended if we include a finite piece in $\Wct$. In the derivation of Eqs.~\eqref{appeq:AWct} and \eqref{appeq:EWct} we assumed that the counterterms are in minimal subtraction  form
\begin{equation}
\Wct(d)=\frac{1}{d-4}\int\dd^dx\,A(d)\;.
\end{equation}
We may parametrise  finite  counterterms by
\begin{equation}
\Wct'(d)=\Wct(d)+\int\dd^4x\,B(4)\;,
\end{equation}
leading to 
\begin{equation}
\Wren'(d)=\Weff(d)+\Wct'(d)\;,
\end{equation}
where  the effective action is defined from 
\begin{equation}
\Weff(d)=-\log\int\DP\phi\,e^{-S[\phi]}\;.
\end{equation}
Firstly, the addition of the 4-dimensional term $B(4)$ only amounts to introducing a classical (i.e independent of $\phi$) term in the UV action
\begin{equation}
\Wren'(d)=\Weff'(d)+\Wct(d)\;,
\end{equation}
with
\begin{equation}
\Weff'(d)=-\log\int\DP\phi\,e^{-S[\phi]-\int\dd^4x\,B(4)}\;,
\end{equation}
hence cannot alter the anomaly, even if it is not Weyl invariant (it would then just be an explicit breaking).

If we were to proceed with the counterterms $\Wct'$ instead of $\Wct$, the explicit breaking and anomaly formulae would need to be amended to
\begin{equation*}
 \mathcal{E}_{\mathrm{Weyl}}  \equiv \langle\tensor{T}{^\mu_\mu}\rangle_{\text{fin}}+\frac{1}{\g}g_{\mu\nu}\frac{\delta}{\delta g_{\mu\nu}}\int\dd^4x\,B(4) \;, \quad
 \mathcal{A}_{\mathrm{Weyl}} =  \frac{1}{\g}\left[g^{(d)}_{\mu\nu}\frac{\delta }{\delta g_{\mu\nu}}\frac{1}{d-4}\int\dd^dx\,A(d)\right]_{\text{fin}} \;,
\end{equation*} 
which are exactly the same as in Eqs.\,\eqref{appeq:AWct} and \eqref{appeq:EWct}. The anomaly given in terms of the EMT \eqref{eqapp:AETM} is unchanged whether we use $\vev{T^{\mu\nu}}=\frac{1}{\g}\frac{\delta \Weff}{\delta g_{\mu\nu}}$ or $\vev{T'^{\mu\nu}}=\frac{1}{\g}\frac{\delta \Weff'}{\delta g_{\mu\nu}}$.

The irrevelance of finite terms of $\Wct$ regarding the anomaly is the reason why we may use the tensor
\begin{equation}
C(d)=R_{\mu\nu\rho\sig}R^{\mu\nu\rho\sig}-2 R_{\mu\nu}R^{\mu\nu}+\frac{1}{3}R^2\;,
\end{equation}
in \eqref{eq:Wctint}, which reduces to the Weyl tensor squared $C(4)=W^2(4)$ 
in four dimensions, instead of the natural extension of the Weyl tensor squared to $d$ dimensions
\begin{equation}
W^2(d)=R_{\mu\nu\rho\sig}R^{\mu\nu\rho\sig}-\frac{4}{d-2} R_{\mu\nu}R^{\mu\nu}+\frac{2}{(d-2)(d-1)}R^2\;.
\end{equation}
Indeed, this second choice of analytic continuation will only amount to finite terms in $\Wct$ since
\begin{equation}
\frac{1}{d-4}W^2(d)=\frac{1}{d-4}C(d)+\mathcal{O}(1)\;.
\end{equation}

\paragraph{Explicit breaking counterterms}
Let us prove a relation that we use regarding the contribution from $\Wct$ that cancels the divergence of the explicit breaking.
Since \eqref{app:deltaWren2} is finite we have
\begin{equation}\label{eq:Tmumudiv1}
\g\langle\tensor{T}{^\mu_\mu}\rangle_{\text{div}}=- \left[g^{(d)}_{\mu\nu}\frac{\delta \Wct(d)}{\delta g_{\mu\nu}}\right]_{\text{div}}\;.
\end{equation}
On the other hand, we also use \eqref{eq:explicitbreakingct}
\begin{equation}
g^{(4)}_{\mu\nu}\frac{\delta \Wct(d)}{\delta g_{\mu\nu}}  = - \g\langle\tensor{T}{^\mu_\mu}\rangle_\Div\;,
\end{equation}
in the derivation of \eqref{eq:genacc}. This relation follows from the fact that $g^{(4)}_{\mu\nu}\frac{\delta \Wct(d)}{\delta g_{\mu\nu}}=0$ in a classically Weyl invariant theory. Therefore, in an explicitly broken theory this is exactly the contribution from $\Wct$ that cancels the divergence of the explicit breaking $\vev{\TEMT}$.

When combined, these last two equations imply
\begin{equation}\label{eq:Wctrelation}
g^{(4)}_{\mu\nu}\frac{\delta \Wct(d)}{\delta g_{\mu\nu}}=\left[g^{(d)}_{\mu\nu}\frac{\delta \Wct}{\delta g_{\mu\nu}}\right]_{\text{div}}\;.
\end{equation}
Let us show this equality explicitly.
The most generic form of the counterterms, including again only $P$-even operators, is
\begin{equation}
\Wct(d)=\frac{1}{d-4}\int\dd^d x\, \g \,\left(a\,E(d) +b\,R(d)^2+ c\, W(d)^2 + f\,\tr\,F(d)^2\right)\;,
\end{equation}
where the $R^2$-term is introduced since Weyl symmetry is explicitly broken.
 As shown above, in a classically Weyl invariant theory we have $g^{(4)}_{\mu\nu}\frac{\delta \Wct(d)}{\delta g_{\mu\nu}}=0$ \eqref{app:Wctvariation2} and  $g^{(d)}_{\mu\nu}\frac{\delta \Wct}{\delta g_{\mu\nu}}$ is finite \eqref{app:Wctvariation1}. Hence, only the $R^2$-term may contribute in \eqref{eq:Wctrelation}. Let us consider its variation under a generic metric variation $\delta g_{\mu\nu}$
\begin{equation}\label{app:deltaR2}
\frac{1}{\g}\delta\,\g\,R^2=\frac{1}{2} g^{\mu\nu}R^2\delta g_{\mu\nu}-2R\,R^{\mu\nu}\delta g_{\mu\nu}+2\,R\, g^{\mu\nu}\left(\nabla_\al \delta\Gamma^\al_{\mu\nu}-\nabla_\mu\delta\Gamma^\al_{\nu\al}\right)\;,
\end{equation}
where
\begin{equation}
\delta\Gamma^\sig_{\mu\nu}=\frac{1}{2}g^{\sig\lambda}\left(\nabla_\nu \delta g_{\mu\lambda}+\nabla_\mu \delta g_{\nu\lambda}-\nabla_\lambda \delta g_{\mu\nu}\right)\;.
\end{equation}
In order to perform the metric differentiation the last term of \eqref{app:deltaR2} is integrated by parts, 
using $\delta g_{\mu\nu}(x)\underset{x\to\infty}{\longrightarrow}0$. We obtain
\begin{equation}
\frac{\delta}{\delta g_{\mu\nu}}\int\dd^dx\,\g\,R^2=\frac{1}{2}\g\,R\left(g^{\mu\nu}R-4R^{\mu\nu}\right)+2\g\,\left(D^\mu D^\nu R-g^{\mu\nu}\Box R\right)\;.
\end{equation}
Finally, using  $g^{(d)}_{\mu\nu}g^{(d),\mu\nu}=d$ and  $g^{(4)}_{\mu\nu}g^{(d),\mu\nu}=4$, as well as \eqref{app:Wctvariation1} and \eqref{app:Wctvariation2} we obtain
\begin{alignat}{2}
&\frac{1}{\g}g^{(4)}_{\mu\nu}\frac{\delta \Wct(d)}{\delta g_{\mu\nu}}&\;=\;&-\frac{6\,b}{d-4}\Box R \nonumber\\
&\frac{1}{\g}g^{(d)}_{\mu\nu}\frac{\delta \Wct(d)}{\delta g_{\mu\nu}}&\;=\;&X(a,c,f)+\frac{b}{2}R^2-2\,b\,\Box R -\frac{6\,b}{d-4}\Box R\;,
\end{alignat}
where $X(a,c,f)$ denotes the contributions from the $E$, $C$ and $F^2$ terms which are finite \eqref{app:Wctvariation1}. As expected, we confirm \eqref{eq:Wctrelation}
\begin{equation}
g^{(4)}_{\mu\nu}\frac{\delta \Wct(d)}{\delta g_{\mu\nu}}=-\frac{6\,b}{d-4}\Box R=\left[g^{(d)}_{\mu\nu}\frac{\delta \Wct}{\delta g_{\mu\nu}}\right]_{\text{div}}\;.
\end{equation}
This is the contribution from $\Wct$  that cancels the divergence of the explicit breaking $\vev{\TEMT}$. Once again, this relation still holds when $P$-odd operators are included in $\Wct$ since they vanish at the level of $\delta \Wct/\delta g_{\mu\nu}$.

\section{Constraints on the Energy-momentum Tensor}
\label{app:EMTconstraints}

In this appendix, we discuss the basis of $P$-odd operators, and we detail how to enforce the diffeo-Lorentz-gauge constraints.

\subsection{Pure gravity case}
Let us recall the generic ansatz \eqref{eq:Talphabeta}
\begin{align}
\label{eqapp:Talphabeta}
\Tct^{\al\be}=\frac{1}{\g}\frac{\delta \Wct}{\delta g_{\al\be}}=&e\,g^{\al\be}R\tilde R+\frac{1}{d-4}\bigg\{e_1\,g^{\al\be}\RRt+e_2 \,P^{\al\be}+e_3\, Q^{(\al\be)}+e_4 \,S^{(\al\be)}\nonumber\\
& +a_1\,g^{\alpha\beta}R^2+a_2\,g^{\alpha\beta}R_{\mu\nu}R^{\mu\nu}+a_3\,g^{\alpha\beta}R_{\mu\nu\rho\sigma}R^{\mu\nu\rho\sigma}+a_4\,g^{\alpha\beta}\Box R\nonumber\\
&+b_1\,R R^{\alpha\beta}+b_2\, R^{\alpha\lambda}\tensor{R}{^\beta_\lambda}+b_3\,R_{\mu\nu}\tensor{R}{^\mu^\alpha^\nu^\beta}+b_4\,R^{\alpha\lambda\mu\nu}\tensor{R}{^\beta_\lambda_\mu_\nu}\nonumber\\
&+c_1\, D^{\alpha} D^{\beta}R+c_2\,\Box R^{\alpha\beta}\bigg\}\;,
\end{align}
with the symmetrisation of indices $t^{(\mu\nu)}=\frac{1}{2}(t^{\mu\nu}+t^{\nu\mu})$\;. The $P$-odd coefficients are constrained by \eqref{eq:deltaLodd4}
\begin{equation}
2\,e_1+e_2+e_3=0\;.
\end{equation}

\subsubsection{Parity-odd operators basis}
\label{app:podd}

Let us try to write all possible parity-odd 2-tensors of mass dimension four depending only on  curvature invariants. 
Firstly, by enumeration it is possible to show that there exists no $P$-even antisymmetric 2-tensor, therefore we have $\epsilon^{\al\be\mu\nu}\mathcal{O}_{\mu\nu}=0$ where $\mathcal{O}$ is a $P$-even 2-tensor. Secondly, due to the Bianchi identities one has 
$\epsilon^{\al\nu\rho\sig}R_{\be\nu\rho\sig}=0$ and $\epsilon^{\mu\nu\rho\sig}D_\nu R_{\al\be\rho\sig}=0$. 
Using this result, any operator can be related to these by Bianchi identities and algebra
\begin{alignat}{2}
&g_{\al\be}R\tilde R&\;=\;&g_{\al\be}\frac{1}{2}\epsilon^{\mu\nu\rho\sigma}R_{\gamma\delta\mu\nu}\tensor{R}{^\gamma^\delta_\rho_\sigma} \;, \\[0.1cm]
&P_{\alpha\beta}&\;=\;&\epsilon^{\mu\nu\rho\sigma}R_{\alpha\lambda\mu\nu}\tensor{R}{_\beta^\lambda_\rho_\sigma}\nonumber\;, \\[0.1cm]
&Q_{\alpha\beta}&\;=\;&\tensor{\epsilon}{_\alpha^\nu^\rho^\sigma}R_{\beta\nu\gamma\delta}\tensor{R}{_\rho_\sigma^\gamma^\delta}\nonumber\;, \\[0.1cm]
&S_{\alpha\beta}&\;=\;&\tensor{\epsilon}{_\alpha^\nu^\rho^\sigma}R_{\beta\lambda\rho\sigma}\tensor{R}{_\nu^\lambda}=-\tensor{\epsilon}{_\al^\nu^\rho^\sig}D_\nu D_\gamma \tensor{R}{_\be^\gamma_\rho_\sig}\nonumber\, .
\end{alignat}
However, upon using the intrinsically four-dimensional Schouten identity  \cite{Remiddi:2013joa}, we obtain
\begin{equation}\label{eqapp:Schouten}
P_{\al\be} = Q_{\al\be}=\frac{1}{2}g_{\al\be} R\tilde R\;,\quad\quad S_{\al\be} = 0\;,
\end{equation}
showing that these operators are not independent in $d=4$. These relations can also directly be obtained from Ref.\cite{Chung:2022ees}, or from  useful applications of the Schouten identities (e.g. \cite{Chala:2021cgt}).

\subsubsection{Diffeomorphism anomaly constraint}

The finiteness of the diffeo anomaly is ensured by
\begin{equation}
D_\al\Tct^{\al\be}=0\; .
\end{equation}
Using Bianchi identities and algebra, it is possible to write the $P$-even part of the divergence of $\Tct$ in terms of independent operators as
\begin{align}
\left. D_\al\Tct^{\al\be}\right|_{P-\mathrm{even}}&=(a_4+c_1+\frac{1}{2}c_2)\Box D^\beta R+(-a_4+b_1+\frac{1}{2}b_2+\frac{1}{2}c_2)\tensor{R}{^\beta^\mu}D_\mu R\\
&+(2\,a_2+b_3-c_2)R^{\mu\nu}D^\beta R_{\mu\nu}+(2\,a_1+\frac{1}{2}b_1)R D^\beta R\nonumber\\
&+(2\,a_3+\frac{1}{2}b_4)R^{\mu\nu\rho\sigma}D^\beta R_{\mu\nu\rho\sigma}+(b_2-b_3+2\,c_2)R^{\mu\nu}D_\mu \tensor{R}{^\beta_\nu}\nonumber\\
&+(-b_3-2\,b_4+2\,c_2)\tensor{R}{^\beta_\nu_\rho_\sigma}D^\sigma R^{\nu\rho}\nonumber\; .
\end{align}
Note that it is not allowed to use integrations by parts at this level (besides $D_\al\Tct^{\al\be}$ is itself a boundary term). Imposing $\left.D_\al\Tct^{\al\be}\right|_{P-\mathrm{even}}=0$ yields seven constraints on the ten parameters associated to the $P$-even operators. They can be recast  to obtain \eqref{eq:7}.

Concerning the $P$-odd part, it is possible to write the divergence as
\begin{align}\label{eqapp:GravDTodd}
\left. D_\nu\Tct^{\mu\nu}\right|_{P-\mathrm{odd}}&=2\,e\, W^\mu(4)+\frac{1}{d-4}\bigg\{2\,e_1\,W^\mu+e_2\,(2\,V^\mu+W^\mu)\nonumber\\
&+\frac{e_3}{2}(2\,X^\mu+Y^\mu+W^\mu)+\frac{e_4}{2}(X^\mu+V^\mu+2\,Z^\mu)\bigg\}=0\; ,
\end{align}
where
\begin{alignat*}{3}
&V^\mu&\;=\;& \eps^{\al\be\ga\de}\tensor{R}{^\mu^\la_\al_\be}D_\ga R_{\la\de}\;,\quad W^\mu&\;=\;& \eps^{\al\be\ga\de}\tensor{R}{^\om^\la_\ga_\de}D_\om \tensor{R}{^\mu_\la_\al_\be}\;,\\
&X^\mu&\;=\;&\eps^{\mu\la\al\be}\tensor{R}{_\al_\be^\ga^\de}D_\ga R_{\de\la}\;,\quad Y^\mu&\;=\;&\eps^{\mu\la\al\be}\tensor{R}{^\om_\la_\ga_\de}D_\om \tensor{R}{^\ga^\de_\al_\be}\;,\quad Z^\mu=\eps^{\mu\la\al\be}R_{\ga\la}D_\al \tensor{R}{^\ga_\be}\;. 
\end{alignat*}
Some of these operators may be related by the Schouten identity in $d=4$ only.
 They are not related by Bianchi identities either and are thus independent. Additionally, the pole and the finite parts in \eqref{eqapp:GravDTodd} are independent. Contrary to the $P$-even part, the system \eqref{eqapp:GravDTodd} is overconstrained leaving the sole solution
\begin{equation}
e=e_1=e_2=e_3=e_4=0\;.
\end{equation}

\subsection{Gauge sector}

Let us now consider the gauge sector.

\subsubsection{$P$-odd operator basis}
\label{app:Poddgauge}

Keeping in mind, the Bianchi identity $\eps^{\mu\nu\rho\sig}D_\nu F_{\rho\sig}=0$ and that  $[F,F]$-commutator terms vanish for both abelian and non-abelian gauge groups (by trace-cyclicity), it is possible to show by enumeration that the operators
\begin{equation}
g^{\mu\nu}\tr\,F \tilde F\;,\quad\quad \tr\,\tensor{F}{^(^\mu_\lambda}\tilde F^{\nu)\lambda}\;,
\end{equation}
define a basis of $P$-odd symmetric gauge field dependent 2-tensors of mass dimension four. In $d=4$, the Schouten identity reduces them to one operator
\begin{equation}
\tr\,\tensor{F}{^(^\mu_\lambda}\tilde F^{\nu)\lambda}=\frac{1}{4}g^{\mu\nu}\tr\,F\tilde F\;.
\end{equation}
Without loss of generality, we can thus write
\begin{equation}
\Tct^{\al\be}=\frac{1}{\g}\frac{\delta \Wctodd^{\mathrm{gauge}}}{\delta g_{\al\be}}=h\,g^{\mu\nu}\tr\,F \tilde F+\frac{1}{d-4}\left(h_1\,g^{\mu\nu}\tr\,F \tilde F+h_2\,\tr\,\tensor{F}{^(^\mu_\lambda}\tilde F^{\nu)\lambda}\right) \;.
\end{equation}
Mixed gravity-gauge symmetric 2-tensors can be ignored since they all vanish under the trace, as mentionned in \SEC\ref{sec:CG}.
This is the case since  there exists no mixed gravity-gauge scalar of mass dimension four. For example, for an abelian gauge group we have the symmetric 2-tensors $\epsilon^{\al\nu\rho\sig}\tensor{R}{^\be_\lambda_\rho_\sig}\,\tensor{F}{^\lambda_\nu}+(\al\leftrightarrow\be)$ and $\epsilon^{\al\nu\rho\sig}\tensor{R}{^\be_\nu}\,\tensor{F}{_\rho_\sig}+(\al\leftrightarrow\be)$. They however vanish (by Bianchi for the first one) when contracted with $g_{\al\be}$. 
 
As emphasised in \SEC\ref{sec:CG}, all the $P$-odd 1-tensors of mass dimension three vanish by Bianchi identity, such that
\begin{equation}
\mathcal{H}_{\text{ct}}^{i,\al}=\frac{1}{\g}\frac{\delta \Wctodd^{\mathrm{gauge}}}{\delta A^i_\al}=0\;.
\end{equation}

\subsubsection{Diffeo, Lorentz and gauge anomaly constraints}
The diffeo, Lorentz and gauge constraints can be expressed in terms of independent operators
\begin{equation}\label{eqapp:gaugeEq}
D_\nu\Tct^{\mu\nu}+i\left(\tensor{F}{_\mu^\nu}\right)^i \mathcal{H}_{\text{ct}}^{i,\nu}=h\,\tr\,D^\mu F \tilde F+\frac{1}{d-4}\left(h_1 v^\mu+h_2\left(\frac{1}{4}v^\mu+w^\mu\right)\right)=0\;,
\end{equation}
where
\begin{equation}
v^\mu = \eps^{\al\be\ga\de}\tr\,F_{\al\be}D^\mu F_{\ga\de} \;,  \quad 
w^\mu = \frac{1}{2}\eps^{\mu\la\al\be}\tr\,D_\nu( \tensor{F}{^\nu_\la}F_{\al\be}) \;.
\end{equation}
These operators are related by the Schouten identity in $d=4$ only, otherwise they are independent. Since the pole and finite part are independent as well in \eqref{eqapp:gaugeEq}, the only solution is
\begin{equation}
h=h_1=h_2=0\;.
\end{equation}

\bibliographystyle{JHEP}
\bibliography{biblio}% Produces the bibliography via BibTeX.

\end{document}